\documentclass[acmsmall]{acmart}

\def\BibTeX{{\rm B\kern-.05em{\sc i\kern-.025em b}\kern-.08emT\kern-.1667em\lower.7ex\hbox{E}\kern-.125emX}}

\copyrightyear{2018}
\acmYear{2018}
\setcopyright{acmlicensed}
\acmConference[Woodstock '18]{Woodstock '18: ACM Symposium on Neural Gaze Detection}{June 03--05, 2018}{Woodstock, NY}
\acmBooktitle{Woodstock '18: ACM Symposium on Neural Gaze Detection, June 03--05, 2018, Woodstock, NY}
\acmPrice{15.00}
\acmDOI{10.1145/1122445.1122456}
\acmISBN{978-1-4503-9999-9/18/06}

\usepackage{mathtools}
\usepackage{amsmath}
\usepackage{color}
\usepackage{graphicx}
\usepackage{threeparttable}
\usepackage[linesnumbered, ruled, vlined]{algorithm2e}
\usepackage{amsthm}

\newtheorem{property}{\textbf{Property}}

\definecolor{azure(colorwheel)}{rgb}{0.0, 0.5, 1.0}
\definecolor{frenchblue}{rgb}{0.0, 0.45, 0.73}
\definecolor{bittersweet}{rgb}{1.0, 0.44, 0.37}
\definecolor{green(pigment)}{rgb}{0.0, 0.65, 0.31}
\definecolor{navyblue}{rgb}{0.0, 0.0, 0.5}
\definecolor{purple(html/css)}{rgb}{0.5, 0.0, 0.5}

\usepackage[noend]{algpseudocode}
\usepackage{indentfirst}

\makeatletter
\newcommand{\rmnum}[1]{\romannumeral #1}
\newcommand{\Rmnum}[1]{\expandafter\@slowromancap\romannumeral #1@}
\makeatother
\usepackage{array}
\newcolumntype{C}[1]{>{\centering\arraybackslash}m{#1}}

\setcopyright{none}
\renewcommand\footnotetextcopyrightpermission[1]{}
\begin{document}

%
% The "title" command has an optional parameter, allowing the author to define a "short title" to be used in page headers.
\title[Delay and Price Differentiation in Cloud Computing]{Delay and Price Differentiation in Cloud Computing: A Service Model, Supporting Architectures, and Performance}

%
% The "author" command and its associated commands are used to define the authors and their affiliations.
% Of note is the shared affiliation of the first two authors, and the "authornote" and "authornotemark" commands
% used to denote shared contribution to the research.
%##############################################
\author{Xiaohu Wu}
%\authornote{Work done while with Fondazione Bruno Kessler, Trento, Italy.}%
\affiliation{%
  \institution{Nanyang Technological University}
  \country{Singapore}
}
\email{xiaohu.wu@ntu.edu.sg}

\author{Francesco De Pellegrini}
\affiliation{%
  \institution{University of Avignon}
  \country{France}
  }
\email{francesco.de-pellegrini@univ-avignon.fr}

\author{Giuliano Casale}
\affiliation{%
 \institution{Imperial College London}
 \country{United Kingdom}}
\email{g.casale@imperial.ac.uk}
%#############################################

%
% By default, the full list of authors will be used in the page headers. Often, this list is too long, and will overlap
% other information printed in the page headers. This command allows the author to define a more concise list
% of authors' names for this purpose.
%\renewcommand{\shortauthors}{Wu et al.}
\renewcommand{\shortauthors}{Xiaohu Wu, Francesco De Pellegrini and Giuliano Casale}

%
% The abstract is a short summary of the work to be presented in the article.
\begin{abstract}
Many cloud service providers (CSPs) provide on-demand service at a price with a small delay. We propose a QoS-differentiated model where multiple SLAs deliver both on-demand service for latency-critical users and delayed services for delay-tolerant users at lower prices. Two architectures are considered to fulfill SLAs. The first is based on priority queues. The second simply separates servers into multiple modules, each for one SLA. As an ecosystem, we show that the proposed framework is dominant-strategy incentive compatible. Although the first architecture appears more prevalent in the literature, we prove the superiority of the second architecture, under which we further leverage queueing theory to determine the optimal SLA delays and prices. Finally, the viability of the proposed framework is validated through numerical comparison with the on-demand service and it exhibits a revenue improvement in excess of 200\%. Our results can help CSPs design optimal delay-differentiated services and choose appropriate serving architectures.
\end{abstract}


%
% Keywords. The author(s) should pick words that accurately describe the work being
% presented. Separate the keywords with commas.
\keywords{QoS-differentiation, incentive compatible, cloud computing}



\maketitle

%\IEEEpubidadjcol

%%%%%%%%%%%%%%%%%%%%%%%%%%%%%%%%%%%%%%%%%%%%%%%%%%%%%%%%%%%%%%%%%%%%%%%%%%%%%%%%%%%%%%%%%%%%%%%%%%%%
%%%%%%%%%%%%%%%%%%%%%%%%%%%%%%%%%%%%%%%%%%%%%%%%%%%%%%%%%%%%
\section{Introduction}%
\label{sec.Introduction}
%%%%%%%%%%%%%%%%%%%%%%%%%%%%%%%%%%%%%%%%%%%%%%%%%%%%%%%%%%%%
%%%%%%%%%%%%%%%%%%%%%%%%%%%%%%%%%%%%%%%%%%%sec.motivation%%%%%%%%%%%%%%%%%%%%%%%%%%%%%%%%%%%%%%%%%%%%%%%%%%%%%%%%%

%%%%%%%%%%%%%%%%%%%%%%%%%%%%%%%%%%%%%%%%%%%%%%%%%%%%%%%%%%%%
%\subsection{Motivation}\label{sec.motivation}
%%%%%%%%%%%%%%%%%%%%%%%%%%%%%%%%%%%%%%%%%%%%%%%%%%%%%%%%%%%%

%\subsection{Background and Motivation}

The Infrastructure-as-a-Service (IaaS) market is projected to grow to \$61.9  billion in 2021 from \$30.5 billion in 2018~\cite{gartner-rp-1}, and is attracting users with different purposes to run their applications on cloud servers. Many cloud service providers (CSPs) provide the standard on-demand service, which is always available at a publicly known price $p$ with a small delay. When a customer\footnote{In this paper, we use customers and users interchangeably.} arrives, it requests to occupy servers for some period without interruption, and a delay arises, i.e., the time from the request arrival to the service commencement. While delay is a key constraint to resource efficiency, customers often differ in the sensitivity to it \cite{Zhou16,amazon-spot-users}. Price differentiation by delays is thus an important research direction to satisfy the customer' preference. Related schemes often use queuing theory for performance analysis and incentive compatibility (IC) to ensure user truthfulness, eliminating the unpredictable effect of non-truthful strategic behaviour on the performance.
%{\color{blue}The related schemes often use queuing theory for performance analysis and the notion of incentive compatibility (IC) to make users truthfully reveal their private information, eliminating the unpredictable effect of non-truthful strategic behaviour on the performance.}

One line of work considers an architecture of separating servers into two parts respectively for on-demand and spot markets \cite{Dierks19,Abhishek12}. Each customer has an initial individual willingness-to-pay (WTP) that further decreases linearly with the delay. The associated slope $c$ defines how sensitive it is to delay, and is called delay-cost type. It will choose to join one market or neither to maximize its surplus. For any customer of spot market, Abhishek {\em et al.} show that, there is a pricing rule to form a Bayesian-Nash incentive compatible mechanism (BNIC), i.e., it will truthfully bid $c$ if the others also do so \cite{Abhishek12}. The customers of higher bids can preempt the servers of others, and each type of customers has an individual service class whose delay relies on the job arrival rate of higher bids. Dierks and Seuken extend the model by considering additional constraints such as preemption cost and the capacity finiteness of on-demand market \cite{Dierks19}.

%{\color{blue}Differently from \cite{Dierks19,Abhishek12}, we consider the following dimensions.}

Differently from \cite{Dierks19,Abhishek12}, we consider the following dimensions. First, it is general to use a family of concave functions to more precisely characterize the WTPs of users \cite{Afeche04a,Guerin20a}. Second, in the current cloud markets, the price $p$ of a CSP is a predefined value that depends on not only WTPs but also other factors such as competition. It is acceptable for most users.
%some delay-tolerant users can also accept delayed services at a lower price.
We thus consider the case that the initial WTPs of users are all $p$, implying their acceptability of on-demand service, and study price differentiation by delays under such context. Third, the delay-cost types can be tremendous and it is operationally costly to maintain an individual service level agreement (SLA) for each type of users \cite{Garg14a}. Fourth, we focus on non-preemptive scheduling, i.e., the service is continuously provisioned to the customer with no interruption. Preemptions are costly and can increase uncertainties  within the delays \cite{Psychas17a,Dargie14a}.

%\subsection{Our Contribution}

\vspace{0.22em}\noindent\textbf{Model.} The standard on-demand service is the fastest service and designed with the principle of ``one size fits all'' to satisfy all types of users. We propose a model of offering a limited number of SLAs to provide incentives and service differentiation among users. These SLAs include both on-demand service for latency-critical jobs, and services with different levels of delay at lower prices.
%All customers' initial WTPs are $p$, implying their acceptability of on-demand service, and characterized by a family of concave functions \cite{Afeche04a}.
%Every user reports its delay-cost type and the CSP will assign a SLA that maximizes its surplus.
The service model is supported by an underlying architecture to fulfill SLAs. Two typical ones are considered. One is similar to the spot market in \cite{Dierks19,Abhishek12}, called the priority-based sharing (PBS) architecture, where delay-tolerant jobs can access the servers of on-demand market in lower priorities. The other simply separates servers into multiple modules, each for one SLA, called the separated multi-SLAs (SMS) architecture.

The proposed model may benefit all market participants. Potential customers get opportunity to trade their delay tolerance for cheaper service. The CSP can thus attract more such customers from its competitors. In queuing systems, the larger the delay, the higher the resource utilization. Delay-differentiated services allow processing more workload than a pure on-demand service, possibly improving its revenue.

\vspace{0.28em}\noindent\textbf{Results.} As an ecosystem, we derive the main features of the model above, and the main results of this paper are as follows:
\begin{enumerate}

\item [(\rmnum{1})] We derive a generic pricing rule that gives the optimal SLA prices when the SLA delays are given in advance, and show that the proposed model is dominant-strategy incentive compatible (DSIC): every user truthfully reports its delay-cost type, regardless of what the others do. DSIC is a stronger degree of IC than BNIC that assumes that an individual customer has the global knowledge of the distribution on user types, which is not needed in DSIC \cite{Nisan07a}. %Furthermore, there exist consecutive intervals such that the customers whose types are in an interval belong to the same SLA. This helps the CSP and customers understand the segmentation of a delay-differentiated cloud market.

\item [(\rmnum{2})] The architecture determines the model's performance.
%The adopted architecture determines the performance of the proposed service model, as well as the SLA delays and market segmentation.
We derive two performance bounds respectively under the PBS and SMS architectures. They show the superiority of a SMS-based service system where a PBS-based system in fact achieves a similar revenue to a pure on-demand system. We then leverage queueing theory to give the optimal SLA prices and delays of a SMS-based system. Finally, we give numerical results to show that it can significantly outperform the standard on-demand service model, with a revenue improvement in excess of 200\%.
\end{enumerate}

The rest of this paper is organized as follows. In Section~\ref{sec.related-works}, we introduce the related work. We propose the delay-differentiated service model in Section~\ref{sec.model}. Next, we study in Section~\ref{sec.strategyproof-mechanism} the related pricing problems. We describe two architectures in Section~\ref{sec.serving-architecture} to support the service model differently, and analyze their performance and optimal parameter configuration. Simulations are done in Section~\ref{sec.performance-evaluation} to evaluate the performance numerically. Finally, we conclude this paper in Section~\ref{sec.conclusion}. Due to space limitation, all proofs of conclusions are put in the appendix.

\section{Related Work}
\label{sec.related-works}

CSPs can offer spot service where customers bid to utilize servers, similar to what Amazon Elastic Cloud Compute (EC2) does. The combination of queue and game theories is used to characterize user behavior and request serving \cite{Casale14a,Adam17a}. %The combination of priority queue and game theory has been used to analyze the performance of two different service models for this purpose \cite{Abhishek12,Dierks19,Wu19a}.
Currently, two main models exist for CPS systems.

{\em The first model} is proposed by Abhishek {\em et al.} \cite{Abhishek12}, which has been partly introduced before. There are $n$ classes of jobs whose mean service time is $s$. Each job of type $i\in [1, n]$ has an initial WTP $v_{i}$ and a linear delay-cost type $C_{i}$ that is a random variable in $[0,\, s\cdot v_{i}]$. The on-demand market is modeled as a $G/G/\infty$ queue with infinite servers to guarantee that the service delay $\varphi$ is zero for all jobs. The spot market is modeled as a preemptive $G/G/m$ queue with finite servers in which each job bids; the higher its bid, the higher its priority to access servers and the lower its delay. Dierks and Seuken extend the first model by considering additional constraints and modeling the on-demand market as a $G/G/m$ queue with finite servers \cite{Dierks19}; thus, the on-demand service is delivered to customers with a small delay $T$. Finally, their mathematical expressions are instantiated with regard to $M/M/m$ queues, and numerical results are given to show the concrete revenue improvement of this model over the pure on-demand service model.

{\em The second model} focuses on enabling users to utilize the idleness of on-demand market. The idle periods of servers appear at random and are utilized as spot service by users who bid the highest \cite{Devanur17}. Wu {\em et al.} show that the challenge is guaranteeing the immediacy of on-demand service and the persistence of spot service while sharing servers \cite{Wu19a}. Then, they give an integral resource allocation and pricing framework for this purpose, and it forms a DSIC mechanism. They basically follow the pricing principle used by Amazon EC2 in practice \cite{Ben-Yehuda} and show how to run such services in cloud systems. Song and Gu\'erin focus on the statistical features of the spot pricing aspect and give the optimal pricing and bidding strategies for a CSP and its users respectively \cite{Guerin20a}. The spot service is also delay-differentiated in the sense that if a user bids higher, its delay will be smaller. For the sake of tractability, the authors also use a family of linear delay-cost functions to characterize the users' sensitivity to delay; numerical results are given for the more general settings including a family of concave functions. Finally, spot service is popular in that users can trade their delay tolerance for cheaper service. However, it indeed creates significant complexity that users have to face and does not provide any delay guarantee \cite{Kash16a,Wu19b,Dubois16a}.

Additionally, there are many works that use the theory of auction and mechanism design to explore potential frameworks for selling computing resource that take into account deadlines \cite{Azar15a,Zhou16,Jain15a,Wu15a} or virtual machine configuration \cite{Zhang15b}, under which the availability of resource depends on a customer's bid and is uncertain. However, in practice, it is often desirable to offer an on-demand market as an option of customers such that the computing service is certainly available at a fixed unit price, as we see for most of products and services in real world. This is also one motivation of the models of this paper and \cite{Abhishek12,Dierks19,Wu19a}.

%%%%%%%%%%%%%%%%%%%%%%%%%%%%%%%%%%%%%%%%%%%%%%%%%%%%%%%%%%%%%%%%%%%%%%%%%%%%%%%%%%%%%%%%%%%%%%%%%%%%
%%%%%%%%%%%%%%%%%%%%%%%%%%%%%%%%%%%%%%%%%%%%%%%%%%%%%%%%%%%%
\section{A QoS-Differentiated Service Model}
\label{sec.model}
%%%%%%%%%%%%%%%%%%%%%%%%%%%%%%%%%%%%%%%%%%%%%%%%%%%%%%%%%%%%
%%%%%%%%%%%%%%%%%%%%%%%%%%%%%%%%%%%%%%%%%%%%%%%%%%%%%%%%%%%%%%%%%%%%%%%%%%%%%%%%%%%%%%%%%%%%%%%%%%%%

In this section, we describe the proposed QoS-differentiated service model, and the associated questions to be addressed. The service model is generic and we postpone the description of the ways of fulfilling its SLAs, which will be given after we study the model properties.

%In this section, we describe the proposed QoS-differentiated service model, and the associated questions to be addressed. The service model is generic and we temporarily omit the description of the ways of fulfilling its SLAs. After we study its properties in Section~\ref{sec.strategyproof-mechanism}, the SLA fulfillment in cloud computing will be introduced in Section~\ref{sec.serving-architecture}.

%\subsection{A QoS-Differentiated Service Model}

\subsection{Delay-Cost Curves}
\label{sec.delay-cost-curve}

%\vspace{0.25em}\noindent\textbf{Delay-Cost Curves.}

Each customer $j$ requests at time $a_{j}$ to occupy servers for some time $s_{j}$. We equivalently refer to such a request as a job $j$, $a_{j}$ as arrival time, and $s_{j}$ as service time. Upon arrival, a job may get served with some delay $\varphi$, i.e., it will get served at time $a_{j}+\varphi$; then, the service stops until the job is continuously served for a duration $s_{j}$. The standard on-demand service in cloud markets represents the fastest service to satisfy all users. We use $T$ and $p$ to denote its delay and price, and they are fixed system parameters: $T$ is the minimum delay before a user can get served where $\varphi\geq T$, and $p$ is the maximum price that a user need to pay for service.

In cloud markets, the WTPs of latency-critical jobs drop sharply even if the delay is increased slightly. For delay-tolerant jobs, although they prefer to get service earlier, their WTPs decrease slowly before the delay increases to a threshold, after which their WTPs decrease sharply \cite{Yeo05a}. The situation of both types of jobs is unified and characterized by a family of functions, denoted by $u(\alpha, \varphi)$.

\begin{property}\label{property-1}
The WTP function $u(\alpha, \varphi)$ is assumed to have the following properties where $\alpha$ is a positive real number and $\varphi\in [T, +\infty)$:
\begin{description}
\item [(\rmnum{1}) Normalisation:] for all $\alpha\in\mathcal{R}^{+}$, we have $u(\alpha, T)=p$;
\item [(\rmnum{3}) Non-increasing:] fixing the value of $\alpha$, $u(\alpha,\varphi)$ is decreasing in $\varphi$;
\item [(\rmnum{2}) Monotone Parametrisation:] fixing the value of $\varphi$, $u(\alpha, \varphi)$ is decreasing in $\alpha$ when $\varphi>T$;
\item [(\rmnum{4}) Decreasing speed:] fixing the value of $\varphi$, $\frac{\partial{u}}{\partial{\varphi}}$ is decreasing in $\alpha$.
\end{description}
\end{property}

The number of users is finite. Each user will choose a specific value of $\alpha$ that can best fit its sensitivity to delay, and $\alpha$ is said to be its delay-cost type. The first subproperty implies that, all users can accept on-demand service at a price $p$ since their WTPs are all $p$ when the delay is $T$. The second subproperty means that, the WTP of a user will decrease as the delay $\varphi$ increases. The third subproperty states under the same delay $\varphi$ that, the larger of the value of $\alpha$, the smaller the WTP $u(\alpha, \varphi)$. Thus, when the delay increases from $T$ to a larger $\varphi$, a user of larger $\alpha$ has more value loss and is more sensitive to delay. $\frac{\partial{u}}{\partial{\varphi}}$ represents the slope of the tangent line at a point. The fourth subproperty guarantees that, if a user has a larger $\alpha$, the decreasing speed of its WTP is also larger.

The function instances can be of any form and the conclusions of this paper will hold only if they satisfy Property~\ref{property-1}. In fact, it can be satisfied by many typical functions in pricing literature. Specifically, the WTP functions can be a family of linear functions in \cite{Dierks19,Abhishek12} where the value loss is characterized by $\alpha\cdot \varphi$ when the delay is $\varphi$. More interestingly, they can also be a family of concave functions \cite{Afeche04a} where the value loss is $\alpha\cdot U(\varphi)$, where $U(\varphi)$ is an increasing convex function; then, $u(\alpha, \varphi)=p-\alpha\cdot U(\varphi)$. As discussed before, they can precisely characterize the following phenomenon: the WTP decreases slightly as delay increases before a threshold; then, it decreases significantly.

%with $T=0$: (\rmnum{1}) for the black curves from left to right, we set $\alpha=$ 5, 2.5, $\frac{5}{3}$, $\frac{5}{4}$, 1 respectively where $\beta=3$; (\rmnum{2}) for the dashed magenta curve, we set $\beta=6$ and $\alpha=5$.

For example, the WTP functions can be instantiated as
\begin{align}\label{equa-wtp}
u(\alpha, \varphi)=p\cdot\left(1-(\alpha\cdot(\varphi-T))^{\beta}\right), \enskip \varphi\in [T, +\infty).
\end{align}
where $\beta\geq 2$ and $\beta$ is a fixed parameter. Here, $U(\varphi)=p\cdot (\varphi-T)^{\beta}$, and $u(\alpha, \varphi)=p-\alpha^{\beta}\cdot U(\varphi)$. We use the term $\alpha^{\beta}$, rather than the term $\alpha$ in \cite{Afeche04a}, as the coefficient of $U(\varphi)$ to simplify the subsequent computation of $\varphi_{0}$; However, this only affects the user's choice of the values of $\alpha$ to specify the same relation between WTP and delay: choosing $\alpha^{\prime}$ with the function (\ref{equa-wtp}) is equivalent to choosing $\alpha^{\prime\prime}=(\alpha^{\prime})^{\beta}$ with the function of the form in \cite{Afeche04a}. The function in (\ref{equa-wtp}) is concave and satisfies Property~\ref{property-1}. Given a customer of type $\alpha$, its WTP becomes zero when the experienced delay $\varphi$ equals $\varphi_{0}=\frac{1}{\alpha}+T$, i.e., $u(\alpha, \varphi_{0})=0$. When the customer experiences a delay no smaller than $\frac{1}{\alpha}+T$ (i.e., $\varphi\geq \frac{1}{\alpha}+T$), it will not accept any service since its WTP is not positive.

\begin{figure}[t]
  \centering
  \includegraphics[width=3.15in]{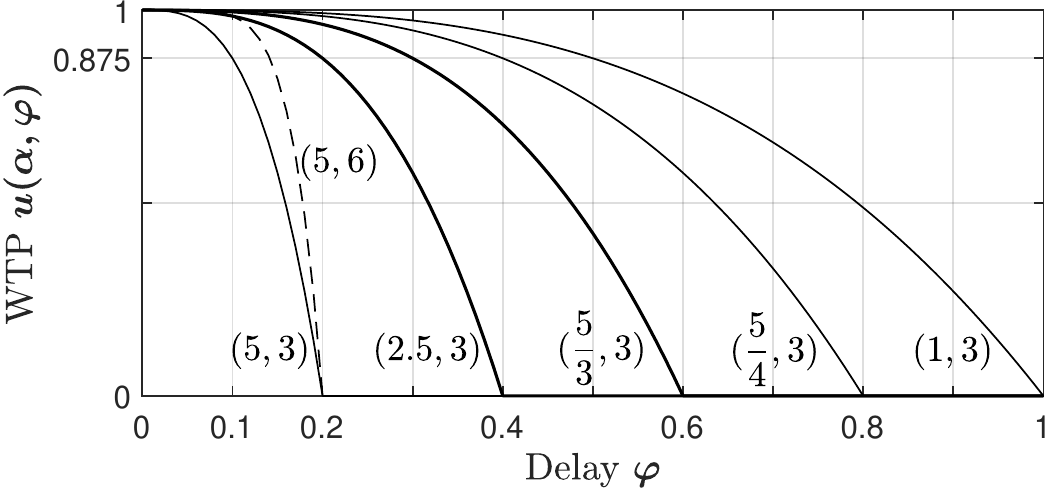}%\\
  \caption{The Curves of WTP Function in (\ref{equa-wtp}) for different values of the parameters $(\alpha, \beta)$.}\label{Fig.1}
\end{figure}

We illustrate the function (\ref{equa-wtp}) in Fig.~\ref{Fig.1} where $T$ is set to zero. As illustrated by the solid curves, latency-critical and delay-tolerant users can respectively choose larger and smaller $\alpha$ to reflect their sensitivities to delay, conforming to the explanation of the third subproperty.
%The larger the value of $\alpha$, the higher its sensitivity to delay. This is illustrated by the solid curves where $\beta=3$.
As illustrated by the leftmost solid curve where $\varphi_{0}=0.2$, a user's WTP will decrease faster and faster as the delay increases, since the function in (\ref{equa-wtp}) is concave. When the delay $\varphi$ ranges in the first half interval $\left[0,\, \frac{\varphi_{0}}{2}\right]$, the WTP decreases slowly from $p$ to $0.875\cdot p$; as $\varphi$ becomes larger and ranges in the second half interval $\left[\frac{\varphi_{0}}{2},\, \varphi_{0}\right]$, the WTP decreases fast from $0.875\cdot p$ to $0$.

Finally, the value of $\beta$ affects the decreasing speed of WTP as the delay increases. The leftmost solid and dashed curves illustrate the cases with $\beta=3$ and $\beta=6$. A larger $\beta$ means that the initial decreasing speed is smaller but then turns larger. In this paper, we study the property of a market that consists of customers whose sensitivities to delay are defined by the values of $\alpha$; when the instance in (\ref{equa-wtp}) is applied, the parameter $\beta$ is common.
%A larger $\beta$ means that the WTP will initially decrease slowly and then decrease to zero quickly. In this paper, we study the property of a market that consists of a type of customers whose sensitivities to delay are defined by the values of $\alpha$. When the instance in (\ref{equa-wtp}) is applied, the parameter $\beta$ is fixed and it dominates the decreasing speed of customers' WTPs.

%The {\color{blue}standard} on-demand service is designed for satisfying all customers {\color{blue}without interruption} in a timely fashion and at a fixed price $p$; the delay of delivering service to customers is guaranteed to be a negligible value, denoted by $T$ \cite{Dierks19}.

\subsection{The QoS-Differentiated Service Model}
\label{sec.qos-differentiated-services}

The CSP plans to offer a finite number of $L$ {\em Service Level Agreements} (SLAs). For all $l\in[1, L]$, the $l$-th SLA specifies a delay $\varphi_{l}$ and the price $p_{l}$ of utilizing a server per unit of time; for the customers operating under the $l$-th SLA, whenever their requests arrive, the CSP guarantees that the expected delay of delivering service is at most $\varphi_{l}$. The first SLA represents the standard on-demand service in cloud markets, and it is for latency-critical users who are not willing to tolerate significant delays. Thus, $p_{1}$ and $\varphi_{1}$ equal the price and delay of an on-demand service. The prices of the other SLAs are lower than $p_{1}$, at the expense of delaying the delivery of computing services to their consumers; here, we let
\begin{align}\label{SLA-delays}
T = \varphi_{1}<\varphi_{2}<\cdots<\varphi_{L}.
\end{align}
Further, we have for all $l\in [1, L-1]$ that the price of the $l$-th SLA is larger than the price of the ($l+1$)-th SLA; otherwise, users would prefer the $l$-th SLA with a smaller delay. Thus, we have $p= p_{1}>p_{2}>\cdots>p_{L}$. We note that $p$ and $T$ are fixed parameters, and $\left\{\varphi_{l}\right\}_{l=2}^{L}$ and $\left\{p_{l}\right\}_{l=2}^{L}$ are decision variables.

\begin{figure}[t]
  \centering
  \includegraphics[width=3.65in]{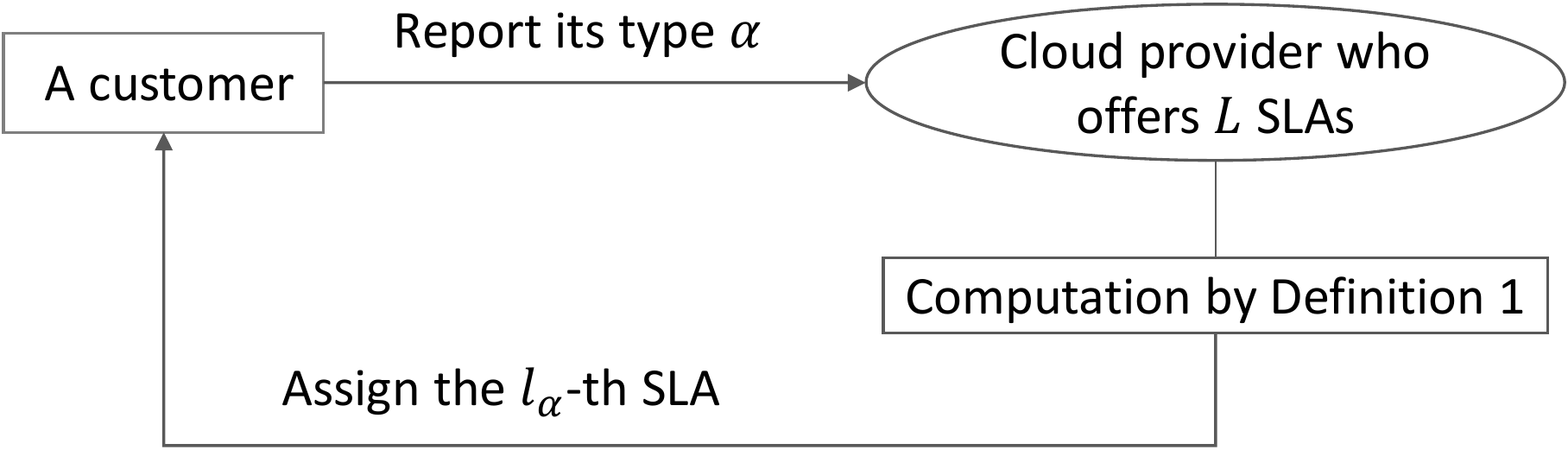}%\\
  \caption{The interaction between customers and a CSP.}\label{Fig.4}
\end{figure}

The interaction process between a CSP and its customers is illustrated in Fig.~\ref{Fig.4}. Specifically, each customer who enters the service system will choose a value $\alpha\in\mathcal{R}^{+}$ such that $u(\alpha, \varphi)$ can best fit its sensitivity to delay; then, it reports the chosen $\alpha$ to the CSP. Users of the same $\alpha$ is said to have the same delay-cost type. The CSP aims to satisfy all its customers, without rejecting any service request, since all customers can accept on-demand service. Under an arbitrary SLA $l\in [1, L]$, the surplus of a customer is its WTP minus the SLA price, i.e., $u(\alpha, \varphi_{l})-p_{l}$. According to the reported type, the CSP will choose one SLA for each type of customers such that their surplus is maximized. Formally, we have the following definition.
%The process of selecting a SLA is illustrated in Fig.~\ref{Fig-SLA-assignment}, and, formally, we have the following definition.

\begin{definition}\label{def-SLA-assignment}
The customers of type $\alpha$ are assigned the $l_{\alpha}$-th SLA defined below:
\begin{align}
l_{\alpha}=\arg\max\limits_{l\in [1, L]}{u(\alpha, \varphi_{l})-p_{l}}.
\end{align}
Specifically, the CSP regulates that, if the customer achieves the same maximum surplus under multiple SLAs, it will be assigned to the SLA whose number is the largest.
\end{definition}

%\begin{figure}[t]
%  \centering
%  \includegraphics[width=3.2in]{Price-variables-cropped}%\\
%  \caption{The process of selecting a SLA for a customer of type $\alpha$.}\label{Fig-SLA-assignment}
%\end{figure}

\subsection{Problem Description}
\label{sec.problem-description}

Each customer submits its delay-cost type $\alpha$ to the CSP that in turn assigns a specific SLA to it. The types of all customers constitute a set $\Phi$; the minimum and maximum values of the elements of $\Phi$ are $\underline{\alpha}$ and $\overline{\alpha}$. Let $P(\alpha)\in (0, 1)$ denote the probability that an arriving customer has a delay-cost type $\alpha$, where $\sum_{\alpha\in\Phi}{P(\alpha)}=1$. The mean arrival rate of the jobs of all types is $\Lambda$, and the mean job size is $s$. For all $l\in [1, L]$, let $\Phi_{l}$ denote the set of the types of the customers who are assigned to the $l$-th SLA, and $\mathcal{P}=\{\Phi_{1}, \Phi_{2}, \cdots, \Phi_{L}\}$ where $\sum_{l=1}^{L}{\Phi_{l}}=\Phi$ and $\Phi_{l_{1}}\cap\Phi_{l_{2}}=\emptyset$ for all $l_{1}, l_{2}\in [1, L]$ with $l_{1}\neq l_{2}$. Thus, the mean job arrival rate of the $l$-th SLA is
\begin{align}\label{equa-arrival-rate-l}
\Lambda_{l}=\Lambda\cdot\sum\nolimits_{\alpha\in\Phi_{l}}{P(\alpha)}.
\end{align}
The total workload of customers that is processed per unit of time under the $l$-th SLA is $w_{l}=\Lambda_{l}\cdot s$. The revenue from the $l$-th SLA per unit of time is $p_{l}\cdot w_{l} = p_{l}\cdot \Lambda_{l}\cdot s$. The total revenue obtained per unit of time is
\begin{align}\label{equa-objective-fun}
G=\sum\nolimits_{l=1}^{L}{p_{l}\cdot w_{l}}=\sum\nolimits_{l=1}^{L}{p_{l}\cdot \Lambda_{l}\cdot s}.
\end{align}
Above, the system input includes $\Phi$, $P(\cdot)$, $\Lambda$, $s$, $m$, $\varphi_{1}$, $p_{1}$ and the decision variables include $\left\{\varphi_{l}\right\}_{l=2}^{L}$, $\left\{p_{l}\right\}_{l=2}^{L}$, and $\mathcal{P}$.

For all $l\in [1, L]$, the $l$-th SLA guarantees that its jobs experiences a delay of at most $\varphi_{l}$. Let $\Theta=(\varphi_{1}, \varphi_{2}, \cdots, \varphi_{L})$. The $\mathcal{P}$ determines the job arrival rate of each SLA by (\ref{equa-arrival-rate-l}). Roughly, in a queuing system, the more the available servers, the smaller the actual experienced delay of serving jobs. When there are $x$ servers and $\mathcal{P}$ is given, the actual delay $t_{l}$ of the jobs of SLA $l\in [1, L]$ is a non-increasing function of $x$. Suppose there are a total of $x=m$ servers for fulfilling all SLAs. Let $T=(t_{1}, t_{2}, \cdots, t_{L})$, and the CSP will provide the minimum number $m$ of servers needed to fulfill SLAs such that
\begin{align}\label{qos-constraint}
T = h(m, \mathcal{P})\leq \Theta.
\end{align}
We leverage queuing theory to characterize the actual delay of each SLA and concretize the function $h(\cdot)$, which enables us to better focus on the overall performance of the proposed model and will be elaborated in Section~\ref{sec.serving-architecture}.

For the service model, we focus on three questions. In the interaction process illustrated in Fig~\ref{Fig.4}, each user needs to report its type information to the CSP. However, this information is private and customers may seek possible ways to maximize their surplus by misreporting their type information. A mechanism is said to be DSIC if a user gains most or at least not less by being truthful, regardless of what the others do \cite{Nisan07a}. In the context of this paper, we have the following definition.

\begin{definition}\label{def-dsic}
Every user of type $\alpha$ will report a type $\alpha^{\prime}$ to the CSP, with the aim to maximize its surplus. Our service framework is said to be DSIC if the user's surplus is maximized when it truthfully reports its type, i.e., $\alpha^{\prime}=\alpha$, no matter whether other users will truthfully do so or not.
\end{definition}

Thus, {\em The first question} is about providing appropriate incentives via pricing SLAs such that our service framework is DSIC.
%{\em Our first question} is about providing appropriate incentives via pricing such that, in the interaction process, users truthfully report their private information (i.e., their types) to the CSP.
{\em The second} is about market segmentation, i.e., how different types of users are grouped together such that each group of users belongs to the same SLA, when the SLA delays are given in advance, and it characterizes the structural property in the mapping of the types to the SLAs (i.e., $\mathcal{P}$). This helps CSP and users better understand the market structure. Then, we will determine the optimal SLA prices  given a specific market segmentation.
%We need to satisfy (\ref{qos-constraint}) to fulfill SLAs.
{\em the third question} is what architecture of servers should be used to satisfy (\ref{qos-constraint}) for fulfilling SLAs. Then, under a particular architecture, we need to leverage queuing theory to optimally determine the market segementation and SLA delays in order to maximize the revenue (\ref{equa-objective-fun}). The main notation used in this paper is summarized in Table~\ref{table}.

\begin{table}[t]
	\centering
	\begin{threeparttable}[t]
       \label{table-notation}
		\caption{Key Notation}
		\begin{tabular}{| C{2.0cm} | p{10.2cm} |}   %p{1.1cm}|p{3.8cm}|}
			\hline
			{\bf Symbol} & {\bf Explanation}\\ \hline
			$L$ & the number of SLAs \\ \hline

            $\varphi_{l}$ & the delay of the $l$-th SLA \\ \hline

            $p_{l}$ & the price of the $l$-th SLA \\ \hline

            $T$ & the delay of on-demand service where $\varphi_{1}=T$ \\ \hline

            $p$ & the price of on-demand service where $p_{1}=p$ \\ \hline

			$m$ & the total number of servers possessed by a CSP \\ \hline	

			$\Lambda$ & the total job arrival rate \\ \hline	
			
			$\lambda_{l}$ & at a single server, the job arrival rate of the $l$-th SLA \\ \hline

			$\hat{\lambda}_{l}$ & at a single server, the total job arrival rate of the first $l$ SLAs \\ \hline		

			$\Phi$ & the set of the types of all customers \\ \hline		
			
            $\overline{\alpha}$ (resp. $\underline{\alpha}$) & the maximum (resp. minimum) type of $\Phi$  \\ \hline			

			$\Phi_{l}$ & the set of the types of the customers who are assigned to the $l$-th SLA \\ \hline		

	    	$\mathcal{P}$ & the set $\{\Phi_{1}$, $\cdots, \Phi_{L}\}$  \\ \hline	

            $\hat{\alpha}_{1}$, $\cdots$, $\hat{\alpha}_{L+1}$ & a division of $\Phi$ used to define $\Phi_{1}, \cdots, \Phi_{L}$ by (\ref{equa-subset-customers}) \\ \hline	

			$t_{l}$ & the actual job delay of the $l$-th SLA \\ \hline			
		\end{tabular}

		\label{table}
	\end{threeparttable}
\end{table}

\section{Market Properties}
\label{sec.strategyproof-mechanism}

In this section, we suppose the SLA delays $\varphi_{1}, \varphi_{2}, \cdots, \varphi_{L}$ are given; then, we show the market segmentation presents a structural property that there exists a sequence $\hat{\alpha}_{1}$, $\hat{\alpha}_{2}$, $\cdots, \hat{\alpha}_{L+1} \in \Phi$ such that for all $l\in [1, L]$ the customers of the types between $\hat{\alpha}_{l}$ and $\hat{\alpha}_{l+1}$ will be assigned to the $l$-th SLA. Further, we derive the optimal SLA prices under which our framework forms a DSIC mechanism while the CSP's revenue is maximized.

\subsection{Market Segmentation}

{If a customer is more sensitive to delay, its WTP will decrease more quickly while facing the same increment in delay. Formally, we have the following relation on the difference of WTPs under two SLAs.}

%Let us consider two arbitrary customers of types $\alpha_{1}$ and $\alpha_{2}$ where $\alpha_{1} > \alpha_{2}$ and $\alpha_{1}, \alpha_{2}\in \Phi$; the customer of type $\alpha_{1}$ is more sensitive to delay as illustrated in Fig.~\ref{Fig.1}.

%For different types of customers, we first compare their speeds at which the WTPs decrease as the delay increases. Let us consider two types of customers $\alpha_{1}$ and $\alpha_{2}$ where $\alpha_{1} > \alpha_{2}$ and $\alpha_{1}, \alpha_{2}\in \Phi$; the former is more sensitive to delay than the latter as illustrated in Fig.~\ref{Fig.1}. Thus, when the delay increases from the same value $\varphi^{\prime}$ to another value $\varphi^{\prime\prime}$, the former has a larger decrement in WTP than the latter. This is formalized as the following lemma.

\begin{lemma}\label{lemma-utility-difference-order}
{Let us consider two arbitrary customers of types $\alpha_{1}$ and $\alpha_{2}$ with $\alpha_{1} > \alpha_{2}$, and two SLAs $k_{1}$ and $k_{2}$ with $k_{1}<k_{2}$. The customer of type $\alpha_{1}$ is more sensitive to delay as explained for Property~\ref{property-1}; the SLA delays satisfy $\varphi_{k_{1}}<\varphi_{k_{2}}$  by (\ref{SLA-delays}).} Then, we have that the difference of the WTPs of the customer of type $\alpha_{1}$ respectively under the $k_{1}$-th and $k_{2}$-th SLAs is larger than its counterpart for the customer of type $\alpha_{2}$, i.e., $$u(\alpha_{1}, \varphi_{k_{1}})-u(\alpha_{1}, \varphi_{k_{2}}) > u(\alpha_{2}, \varphi_{k_{1}})-u(\alpha_{2}, \varphi_{k_{2}}).$$
\end{lemma}
%\begin{proof}
%{\color{blue}Let $\varphi\in [T, +\infty)$. It suffices to prove the conclusion that $g(\varphi)=u(\alpha_{2}, \varphi)-u(\alpha_{1}, \varphi)$ is an increasing function of $\varphi$; then, the lemma holds since $g(\varphi_{k_{2}})>g(\varphi_{k_{1}})$. To prove this, we note that the derivative of $g(\varphi)$ is
%\begin{align*}
%g^{\prime}(\varphi)=\frac{\partial{u(\alpha_{2}, \varphi)}}{\partial{\varphi}}-\frac{\partial{u(\alpha_{1}, \varphi)}}{\partial{\varphi}}.
%\end{align*}
%Since $\alpha_{1}>\alpha_{2}$, we have $g^{\prime}(\varphi)>0$ by the fourth point of Property~\ref{property-1}, and $g(\varphi)$ is increasing.}
%\end{proof}

According to Definition~\ref{def-SLA-assignment}, the CSP will select for each customer a SLA under which its surplus is maximized. Roughly, a customer of larger $\alpha$ is more sensitive to delay and will be assigned to a SLA with a smaller delay, as shown below.

\begin{lemma}\label{lemma-sequential-choice}
Let us consider two customers of types $\alpha_{1}$ and $\alpha_{2}$ where $\alpha_{1} > \alpha_{2}$. If the customers of types $\alpha_{1}$ and $\alpha_{2}$ are respectively assigned to the SLAs $k_{1}$ and $k_{2}$ (i.e., $\alpha_{1}\in\Phi_{k_{1}}$ and $\alpha_{2}\in\Phi_{k_{2}}$), then we have $$k_{1}\leq k_{2},$$ where the SLA delays satisfy $\varphi_{k_{1}}\leq \varphi_{k_{2}}$ by (\ref{SLA-delays}).
\end{lemma}
%\begin{proof}
%We prove this by contradiction. Suppose $k_{2} < k_{1}$ and the SLA delays satisfy $\varphi_{k_{2}}<\varphi_{k_{1}}$. The customer of type $\alpha_{1}$ (resp. $\alpha_{2}$) achieves the maximum surplus under the SLA $k_{1}$ (resp. $k_{2}$), and we thus have
%\begin{align}
% u(\alpha_{1}, \varphi_{k_{1}})-p_{k_{1}} \geq u(\alpha_{1}, \varphi_{k_{2}})-p_{k_{2}} \label{ineq-1}\\
% u(\alpha_{2}, \varphi_{k_{1}})-p_{k_{1}} \leq u(\alpha_{2}, \varphi_{k_{2}})-p_{k_{2}} \label{ineq-2}
%\end{align}
%Multiplying (\ref{ineq-1}) by -1 and adding the resulting inequality to (\ref{ineq-2}), we have $u(\alpha_{2}, \varphi_{k_{1}})-u(\alpha_{1}, \varphi_{k_{1}}) \leq u(\alpha_{2}, \varphi_{k_{2}})-u(\alpha_{1}, \varphi_{k_{2}})$. {\color{blue}However, since $\alpha_{1} > \alpha_{2}$ and $k_{2}<k_{1}$, we have by Lemma~\ref{lemma-utility-difference-order} that $u(\alpha_{1}, \varphi_{k_{2}})-u(\alpha_{1}, \varphi_{k_{1}}) > u(\alpha_{2}, \varphi_{k_{2}})-u(\alpha_{2}, \varphi_{k_{1}})$, which contradicts the previous inequality.}
%\end{proof}

The following proposition characterizes the market segmentation, i.e., the mapping of the types of customers to the SLAs.

\begin{proposition}\label{theo-market-segmentation}
%The customers of different types are segmented as follows:
There exists a sequence $\hat{\alpha}_{1}, \hat{\alpha}_{2}, \cdots, \hat{\alpha}_{L+1} \in \Phi$ such that the $l$-th SLA will be assigned the customers of type $\alpha\in \Phi_{l}$, where $\underline{\alpha}=\hat{\alpha}_{L+1} < \cdots < \hat{\alpha}_{2} < \hat{\alpha}_{1}=\overline{\alpha}$ and $\Phi_{l}$ is a subset of the customer types defined below:
\begin{align}\label{equa-subset-customers}
\Phi_{l}=
\begin{cases}
\Phi \cap \left(\hat{\alpha}_{l+1}, \hat{\alpha}_{l}\right], & \text{ if } l\in [1, L-1],\\
\vspace{0.3em} \Phi \cap \left[\hat{\alpha}_{L+1}, \hat{\alpha}_{L}\right], & \text{ if } l=L.\\
\end{cases}
\end{align}
\end{proposition}
%\begin{proof}
%Each type of customers will be assigned to some SLA, and $\Phi_{l}$ denotes the set of the types of the customers assigned to the $l$-th SLA for all $l\in [1, L]$. Let $\hat{\alpha}_{l}$ denote the maximum type in $\Phi_{l}$ such that only the customers of type $\alpha \leq \hat{\alpha}_{l}$ will possibly be assigned to the $l$-th SLA. For all $l\in [1, L-1]$, when the customers of types $\hat{\alpha}_{l}$ and $\hat{\alpha}_{l+1}$ are respectively assigned the $l$-th and ($l+1$)-th SLAs, we have by Lemma~\ref{lemma-sequential-choice} that $\hat{\alpha}_{l}>\hat{\alpha}_{l+1}$, which can be easily proved by contradiction. A customer of type $\overline{\alpha}$ will be assigned to a SLA whose number is no larger than one (i.e., the first SLA) since $\overline{\alpha}\geq \hat{\alpha}_{1}$. Thus, we have $\hat{\alpha}_{1}=\overline{\alpha}$.

%By Lemma~\ref{lemma-sequential-choice}, we also have that (\rmnum{1}) for all $l\in [1, L-1]$ every customer of type $\alpha\in \left(\hat{\alpha}_{l+1}, \hat{\alpha}_{l}\right]\cap\Phi$ will be assigned to a SLA whose number $l^{\prime}$ is no smaller than $l$ but no larger than $l+1$, and (\rmnum{2}) every customer of type $\alpha\in \left[\underline{\alpha}, \hat{\alpha}_{L}\right]\cap\Phi$ will be assigned to a SLA whose number is no smaller than $L$ since $\alpha\leq \hat{\alpha}_{L}$. In the first case, $\alpha> \hat{\alpha}_{l+1}$ and $\hat{\alpha}_{l+1}$ is the maximum type of $\Phi_{l+1}$; thus $l^{\prime}$ will be smaller than $l+1$ and equal $l$. The proposition thus holds.
%\end{proof}

Proposition~\ref{theo-market-segmentation} shows that, in a delay-differentiated market, the customers are segmented by a sequence $\hat{\alpha}_{1}, \hat{\alpha}_{2}, \cdots, \hat{\alpha}_{L+1}$ such that the customers of type $\alpha\in\Phi_{l}$ will be assigned to the $l$-th SLA.

\subsection{Optimal DSIC Mechanism}

Let us suppose in this subsection we are given a particular market segmentation $\hat{\alpha}_{1}, \hat{\alpha}_{2}, \cdots, \hat{\alpha}_{L+1}$ defined in Proposition~\ref{theo-market-segmentation}. Then, we will derive the corresponding SLA prices $p_{1}, p_{2}, \cdots, p_{L}$ that simultaneously guarantee that (\rmnum{1}) they are optimal to maximize a CSP's revenue, and (\rmnum{2}) our service framework forms a DSIC mechanism.
%\begin{enumerate}
%\item [(\rmnum{1})] they are optimal to maximize a CSP's revenue;
%\item [(\rmnum{2})] our service framework forms a DSIC mechanism.
%\end{enumerate}
%Conversely, with such SLA prices, the market segmentation keeps unchanged, i.e., every customer of type $\alpha\in\Phi_{l}$ will still be assigned to the $l$-th SLA where $\Phi_{l}$ is given in (\ref{equa-subset-customers}).

%We note that, the second point above may be violated if the SLA prices are not properly set according to the particular market segmentation.

First, we give a definition that is used to define SLA prices.

\begin{definition}\label{def-optip}
Let $u_{l}^{-} = u\left(\hat{\alpha}_{l}, \varphi_{l-1}\right) - u\left(\hat{\alpha}_{l}, \varphi_{l}\right)$ for all $l\in [2, L]$ where $u_{l}^{-}$ is the difference of the WTPs of a customer of type $\hat{\alpha}_{l}$ respectively under the ($l-1$)-th and $l$-th SLAs. We define parameter $\hat{p}_{l}$ to be such that,
\begin{enumerate}
\item [(\rmnum{1})] $\hat{p}_{1}=u\left(\hat{\alpha}_{1}, \varphi_{1}\right)=p$, i.e., the price of on-demand instances;

\item [(\rmnum{2})] for all $l\in [2,\, L]$, $\hat{p}_{l}$ is the maximum possible $p_{l}$ that satisfies $p_{l} \leq \hat{p}_{l-1}-u_{l}^{-}$, i.e.,
    \begin{align*}
    \hat{p}_{l} = \hat{p}_{l}\left( \hat{\alpha}_{1}, \cdots, \hat{\alpha}_{l}, \varphi_{1}, \cdots \varphi_{l} \right) & = \hat{p}_{l-1} - u_{l}^{-}
     = \hat{p}_{1} - \sum\nolimits_{l^{\prime}=2}^{l}{u_{l^{\prime}}^{-}}.
    %= \hat{p}_{1} - \sum\nolimits_{l^{\prime}=2}^{l}{\left(u\left(\hat{\alpha}_{l^{\prime}}, \varphi_{l^{\prime}-1}\right) - u\left(\hat{\alpha}_{l^{\prime}}, \varphi_{l^{\prime}}\right)\right)}.
    \end{align*}
\end{enumerate}
\end{definition}

Second, each type of customers is assigned some SLA according to Definition~\ref{def-SLA-assignment}, and we will show that, when the SLA prices $p_{1}$, $p_{2}$, $\cdots$, $p_{L}$ are set to $\hat{p}_{1}$, $\hat{p}_{2}$, $\cdots$, $\hat{p}_{L}$, the market segmentation is still $\hat{\alpha}_{1}, \hat{\alpha}_{2}, \cdots, \hat{\alpha}_{L+1}$, i.e., every customer of type $\alpha\in\Phi_{l}$ is still assigned to the $l$-th SLA where $\Phi_{l}$ is given by (\ref{equa-subset-customers}).

To prove this, we consider the surpluses of a customer of type $\alpha\in\Phi_{l}$ under two adjoining SLAs whose numbers are simultaneously no larger or smaller than $l$. Roughly, its surplus under the SLA whose number is closer to $l$ is always larger than its surplus under the other SLA, as shown below.

\begin{lemma}\label{lemma-inequality}
Suppose the SLA prices $p_{1}, p_{2}, \cdots, p_{L}$ are set to $\hat{p}_{1}$, $\hat{p}_{2}$, $\cdots$, $\hat{p}_{L}$. Let us consider a customer of type $\alpha\in\Phi_{l}$ and a SLA $l^{\prime}$ where $l, l^{\prime}\in [1, L]$ and $\Phi_{l}$ is given by (\ref{equa-subset-customers}). The surplus of this customer is such that (\rmnum{1}) in the case that $l^{\prime}\in [2, l]$, we have
\begin{itemize}
\item if $\alpha=\hat{\alpha}_{l}$ and $l^{\prime}=l$, its surpluses under the $l^{\prime}$-th and ($l^{\prime}-1$)-th SLAs are the same, and
\item otherwise, its surplus under the $l^{\prime}$-th SLA is larger than its surplus under the ($l^{\prime}-1$)-th SLA;
\end{itemize}
and (\rmnum{2}) in the case that $l^{\prime}\in [l, L-1]$, its surplus under the $l^{\prime}$-th SLA is larger than its surplus under the ($l^{\prime}+1$)-th SLA.
%\begin{enumerate}
%\item [(\rmnum{1})] in the case that $l^{\prime}\in [2, l]$, we have
%\begin{itemize}
%\item if $\alpha=\hat{\alpha}_{l}$ and $l^{\prime}=l$, its surpluses under the $l^{\prime}$-th and ($l^{\prime}-1$)-th SLAs are the same, and
%\item otherwise, its surplus under the $l^{\prime}$-th SLA is larger than its surplus under the ($l^{\prime}-1$)-th SLA;
%\end{itemize}
%\item [(\rmnum{2})] in the case that $l^{\prime}\in [l, L-1]$, its surplus under the $l^{\prime}$-th SLA is larger than its surplus under the ($l^{\prime}+1$)-th SLA.
%\end{enumerate}
\end{lemma}

Using the transitiveness of inequalities, we derive the following proposition with Lemma~\ref{lemma-inequality}.

\begin{proposition}\label{theo-SLA-assignment-maximum}
When the SLA prices $p_{1}, p_{2}, \cdots, p_{L}$ are set to $\hat{p}_{1}$, $\hat{p}_{2}$, $\cdots$, $\hat{p}_{L}$, we have for all $l\in [1, L]$ that a customer of type $\alpha\in \Phi_{l}$ will be assigned to the $l$-th SLA where $\Phi_{l}$ is given by (\ref{equa-subset-customers}). In other words, the customer achieves the maximum surplus under the $l$-th SLA.
\end{proposition}
%\begin{proof}
%In the case that $\alpha\neq\hat{\alpha}_{l}$, we have by Lemma~\ref{lemma-inequality} the conclusion that, (\rmnum{1}) for all $l^{\prime}\in [2, l]$, the customer achieves a higher surplus under $l^{\prime}$-th SLA than under the ($l^{\prime}-1$)-th SLA, and (\rmnum{2}) for all $l^{\prime}\in [l, L-1]$, it achieves a higher surplus under the $l^{\prime}$-th SLA than under the ($l^{\prime}+1$)-th SLA; thus, the customer achieves the highest surplus under the $l$-th SLA. In the case that $\alpha=\hat{\alpha}_{l}$, we still have the above conclusion, except that the customer achieves the same surplus under the $l$-th and ($l-1$)-th SLAs when $l^{\prime}\in [2, l]$ and $l^{\prime}=l$. Hence, the customer achieves the maximum surplus under both the $l$-th and ($l-1$)-th SLAs. According to Definition~\ref{def-SLA-assignment}, the proposition holds in both cases.
%\end{proof}

Third, we show that, when the SLA delays $\varphi_{1}, \cdots, \varphi_{L}$ and market segmentation $\hat{\alpha}_{1}, \cdots, \hat{\alpha}_{L+1}$ are arbitrarily given, there is a pricing rule such that the SLA prices are optimal and our framework forms a DSIC mechanism.

\begin{proposition}\label{theo-sequence-1}
When the SLA prices $p_{1}, p_{2}, \cdots, p_{L}$ are set to $\hat{p}_{1}, \hat{p}_{2}, \cdots, \hat{p}_{L}$, we have
\begin{enumerate}
\item [(\rmnum{1})] our service framework forms a DSIC mechanism;
%\item [(\rmnum{2})] for all $l\in [1, L]$, the customers of types $i_{l}, i_{l}+1, \cdots, i_{l+1}-1$ will be assigned to the $l$-th SLA;
\item [(\rmnum{2})] $\hat{p}_{1}, \hat{p}_{2}, \cdots, \hat{p}_{L}$ are the optimal SLA prices.
\end{enumerate}
\end{proposition}
%\begin{proof}
%Let us consider a customer of type $\alpha\in\Phi_{l}$ who reports to the CSP that its type is $\alpha^{\prime}$. No matter what the other users do, we have by Proposition~\ref{theo-SLA-assignment-maximum} that it achieves the maximum surplus under the $l$-th SLA and will be assigned by the CSP to the $l$-th SLA when it truthfully reports its type, i.e., $\alpha^{\prime}=\alpha$. Thus, it cannot gain more by misreporting its type, since misreport can lead to that it is assigned to the $l$-th SLA or the other SLAs. The first point thus holds by Definition~\ref{def-dsic}.

%The objective of our framework is to maximize (\ref{equa-objective-fun}); given the market segmentation $\hat{\alpha}_{1}, \hat{\alpha}_{2}, \cdots, \hat{\alpha}_{L+1}$ defined in Proposition~\ref{theo-market-segmentation}, the job arrival rate of each SLA is fixed by (\ref{equa-arrival-rate-l}) and we have the conclusion that the larger the SLA prices, the larger the value of $AG$. The first SLA's price $p_{1}$ is fixed and equals $p$. In order to guarantee the truthfulness of the customers of type $\alpha\in\Phi_{l}$, a necessary condition is that $u_{i_{l}}(\alpha, \varphi_{l-1}) - p_{l-1} \leq u_{i_{l}}(\alpha,\varphi_{l}) - p_{l}$, for all $l\in [2, L]$. Further, irrespective of the value of $p_{l-1}$, the maximum possible value of $p_{l}$ is $\hat{p}_{l}$ for all $l\in [2, L]$. Thus, the second point holds.
%\end{proof}

\begin{figure*}[t]
  \centering
  \includegraphics[width=4.3in]{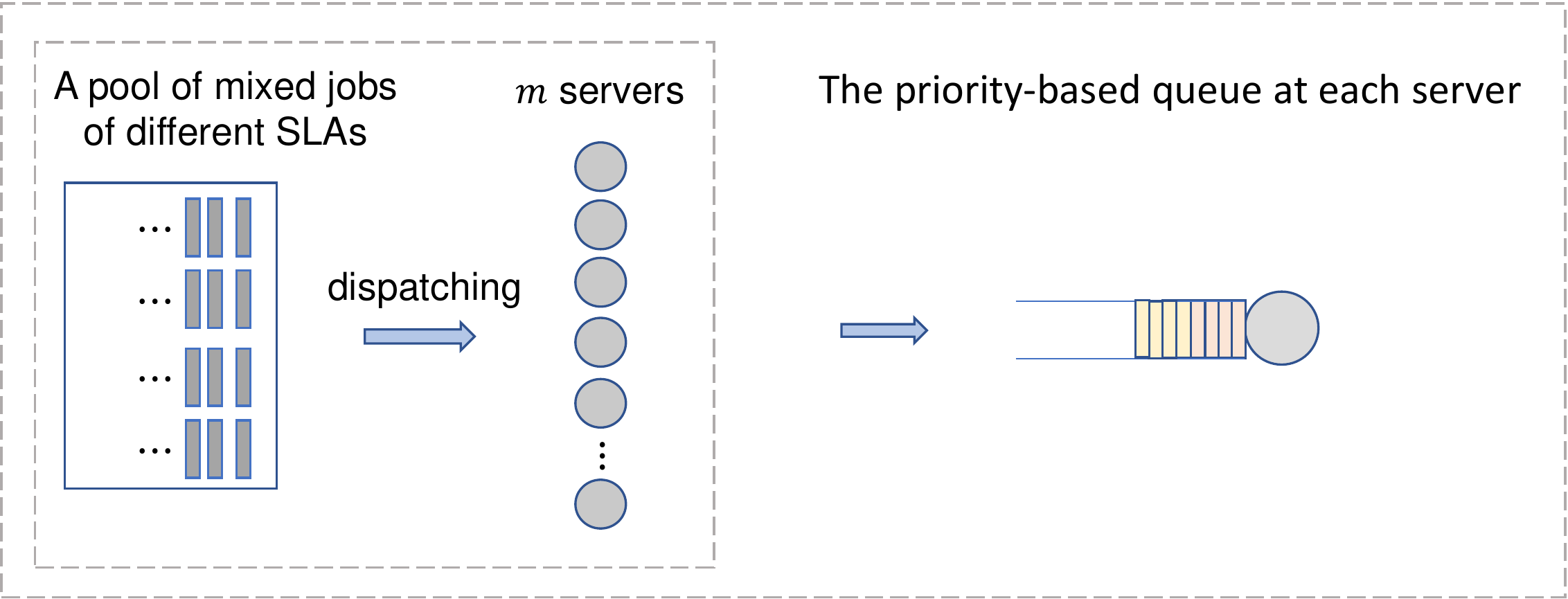}%\\
  \caption{The priority-based sharing architecture with $L=2$: grey rectangles denote all jobs that are dispatched to multiple servers in spite of their SLAs; at a single server, the jobs of the first SLA (denoted by orange rectangles) have a higher priority to be served than the jobs of the second SLA (denoted by golden rectangles).}\label{Fig-priority-based-sharing}
\end{figure*}

%\section{Priority-based Queueing System}
\section{Supporting Architectures, and Their Performance}
\label{sec.serving-architecture}

In Sections~\ref{sec.model} and~\ref{sec.strategyproof-mechanism}, we study a generic service model that offers $L$ SLAs and its properties in pricing and user behavior. The SLA fulfillment relies on proper provision of servers to jobs to satisfy (\ref{qos-constraint}). In this section, we will consider two typical architectures of servers to serve jobs and fulfill SLAs. Then, we study their performance and optimal configuration of parameters such as SLA delays.

\subsection{Two Supporting Architectures}
\label{sec.two-architectures}

A CSP has a total of $m$ servers. When a job $j$ arrives, it is assigned to a server that will serve it for a duration $s_{j}$. We will respectively consider (\rmnum{1}) the PBS architecture and (\rmnum{2}) the SMS architecture. In the former, an arriving job will be assigned to one of the $m$ servers, and the order of serving the jobs at a server depends on their priorities, which depend on the SLAs to which they belong. In the latter, servers are separated into $L$ groups and each exclusively serves the jobs of the same SLA.

\subsubsection{\textbf{Preliminary}}

%Both architectures involve dispatching every arriving job to one of multiple servers.

Before elaborating the architectures, we first introduce the polices used in cloud services for assigning jobs \cite{Esa18}. Suppose there are $m^{\prime}$ servers to serve a particular group of jobs and the mean job arrival rate is $\Lambda^{\prime}$. Typical dispatching policies include (\rmnum{1}) {\em Random}: for every job , it chooses every server with the same probability $\frac{1}{m^{\prime}}$ and assign $j$ to the chosen server \cite{Zheng16a,Rasley16a}, and (\rmnum{2}) {\em Round-Robin (RR)}: jobs are assigned to servers in a cyclical fashion with the $j$-th job being assigned to the $i$-th server where $i=j\, mod\, m^{\prime}$ \cite{Wang14a}.
%\begin{description}
%\item [(\rmnum{1}) Random:] for every job $j$, it chooses a server with the same probability $\frac{1}{m^{\prime}}$ and assign $j$ to this server \cite{Zheng16a,Rasley16a},
%\item [(\rmnum{2}) Round-Robin (RR):] jobs are assigned to servers in a cyclical fashion with the $j$-th job being assigned to the $i$-th server where $i=j\, mod\, m^{\prime}$ \cite{Wang14a,Chung18a}.
%\end{description}
As a result, jobs are evenly dispatched over the $m^{\prime}$ servers. At each server, the arriving jobs form a single queue with the same mean job arrival rate $\lambda^{\prime}=\frac{\Lambda^{\prime}}{m^{\prime}}$ \cite{Esa16a}. The service time of a job is denoted by a random variable $x$ and the mean $s$ of $x$ is normalized to be one, i.e., $s=1$.

The RR and Random policies are also supported by Amazon EC2 to dispatch jobs to servers while on-demand users are being served \cite{Chung18a}. Their prevalence can be due to the following reasons. Each job needs an individual job assignment decision. Such policies do not need the knowledge of server states and can form a distributed scheduler where numerous job assignment decisions could be done instantaneously, thus reducing the scheduling delays. On the other hand, the maintenance of the state information of all servers relies on a heartbeat mechanism where servers communicate on their states with a centralized scheduler at a specific frequency and the job assignment decisions are also made at such a frequency \cite{Karanasos}. In larger-scale server systems like cloud systems, to reduce communication overhead, the frequency has to be low, which leads to a relatively large scheduling delay \cite{Mukherjee,Psychas17a}.

\begin{figure*}[t]
  \centering
  \includegraphics[width=4.8in]{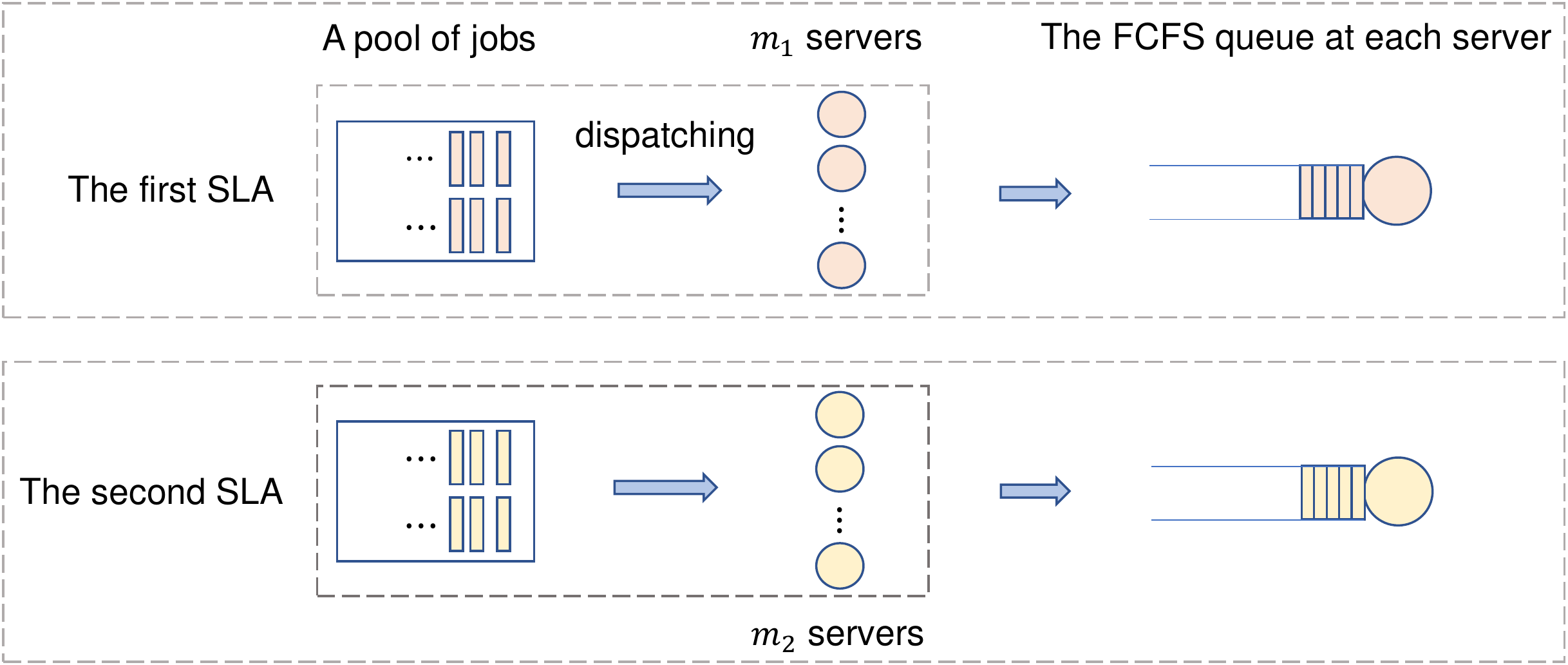}%\\
  \caption{The separated multi-SLAs architecture with $L=2$ and $m_{1}+m_{2}=m$: colored rectangles denote jobs of different SLAs while colored circles denote servers of different SLAs.}\label{Fig-separated-multi-slas}
\end{figure*}

\subsubsection{\textbf{The PBS Architecture}}

In the PBS architecture, whenever a job arrives, it is assigned to one of the $m$ servers by some dispatching policy described above. The total job arrival rate is $\Lambda$ and the job arrival rate at a single server is $\lambda=\frac{\Lambda}{m}$. At every server, the jobs have $L$ priority classes. For all $l\in [1, L-1]$, the jobs of SLA $l$ have higher priority to utilize servers than the jobs of SLAs $l+1$, and are said to have a priority $l$. At the moments of job completion, the server becomes idle and will select a new job of the highest priority to serve, and jobs of the same priority will be chosen in a first-come-first-served (FCFS) discipline. While a job $j$ is being served, the nonpreemptive rule is applied, that is, the job will continuously occupy a server for a duration $s_{j}$ even if other jobs of higher priorities arrive.

%\vspace{0.2em}\noindent\textbf{Job delay.}
Now, we give the mean delay $t_{l}$ of the jobs of each SLA $l\in [1, L]$. At each server, the job arrival rate of the $l$-th SLA is $\lambda_{l} = \lambda\cdot\sum\nolimits_{\alpha\in\Phi_{l}}{P(\alpha)}$. The total arrival rate of the jobs of SLAs $1$,$\cdots$, $l$ is $\hat{\lambda}_{l} = \sum\nolimits_{l^{\prime}=1}^{l}{\lambda_{l^{\prime}}}$. The jobs of all SLAs at every server form a single queue and their job arrivals are described as a Poisson process with rate $\lambda$. The service time $x$ of jobs is assumed to follow a general distribution where the mean $s$ is one. Such a queue is usually denoted by $M/G/1$. We can directly use the result for a $M/G/1$ queue with priority in \cite{Bertsekas87a} and get the delay of the jobs of the $l$-th SLA
\begin{align}\label{equa-waiting-time-random-0}
t_{l} = \frac{0.5\cdot\lambda\cdot E[x^{2}]}{(1-\hat{\lambda}_{l-1})\cdot(1-\hat{\lambda}_{l})},
%t_{l} = \frac{0.5\cdot\lambda\cdot E[x^{2}]}{(1-\hat{\lambda}_{l-1})\cdot(1-\hat{\lambda}_{l})},
\end{align}
where $l\in [1, L]$, $\hat{\lambda}_{0}$ is set to zero trivially, and $E[x^{2}]$ is the second moment of $x$, i.e., its mean-squared
value.

%{\color{blue}For example, when the service time $x$ follows an exponential distribution, we have $E[x^{2}]=2\cdot s^{2}=2$. In the first SLA, $\hat{\lambda}_{1} <1$; the constraint that the delay of the first SLA is no larger than $T$ (i.e., $t_{1}\leq T$) requires that $\lambda<T$. This means that the total load $\lambda\cdot s$ of a server is low and the performance of a PBS-based service system is possibly poor, since $T$ is small. Such observation will be formally elaborated in Section~\ref{Sec.optimal-delay-PBS}.}

\subsubsection{\textbf{The SMS Architecture}}
\label{sec.architecture-sms}

In the SMS architecture, the $m$ servers are separated into $L$ groups, and each group has $m_{i}$ servers and forms a module, where $m=\sum_{l=1}^{L}{m_{l}}$. The $l$-th module is used to exclusively serve the jobs of the $l$-th SLA, and every job that belongs to the $l$-th SLA will be assigned to one of the $m_{l}$ servers under some dispatching policy such as Random or RR. At every server, the jobs will be served in a FCFS discipline. The total job arrival rate of the $l$-th SLA is $\Lambda_{l}$ and the job arrival rate at a single server is $\lambda_{l}=\frac{\Lambda_{l}}{m_{l}}$. The jobs at every server forms a single queue, and when it is a M/G/1 queue, we have from \cite{Bertsekas87a} that the job delay of the $l$-th SLA is
\begin{align}\label{equa-waiting-time-separated-random}
t_{l}= \frac{0.5\cdot \lambda_{l}\cdot E[x^{2}]}{1-\lambda_{l}}.
\end{align}

\subsubsection{\textbf{On-demand Service System}}

The delay-differentiated service system of this paper can be viewed as a complement to the standard on-demand service model, which will be used as a benchmark. In a pure on-demand system, all jobs are served with a short delay and processed with the same priority on the $m$ servers. Upon arrival of each job, it will be dispatched to one of the $m$ servers under some policy and the jobs at the same server will be served in a FCFS discipline. The total job arrival rate is $\Lambda_{od}$, and the job arrival rate at a single server is $\lambda_{od}=\frac{\Lambda_{od}}{m}$. Similar to (\ref{equa-waiting-time-separated-random}), we have that the delay of all jobs is
\begin{align}\label{equa-waiting-time-on-demand}
t= \frac{0.5\cdot \lambda_{od}\cdot E[x^{2}]}{1-\lambda_{od}}.
\end{align}
The job delay will be no larger than $T$, which requires that $t\leq T$.

Beyond the above architectural description, we will use in this paper exponential or hyperexponential distribution to model the service time $x$. As often used in the literature \cite{Mukherjee,Rasley16a}, they have simple closed-form expressions for $E[x^{2}]$ and can guarantee the existence of $E[x^{2}]$, which enable analytically evaluating the performance of the architectures above. When $x$ follows an exponential distribution \cite{Mukherjee,Rasley16a}, we have
\begin{align}\label{equa-exp}
E[x^{2}] = 2\cdot s^{2} = 2.
\end{align}
When $x$ follows a hyperexponential distribution \cite{Mukherjee}, it can be characterized by $h$ tuples $(\pi_{i}, \eta_{i})$ where $i\in [1, h]$ and $\sum_{i=1}^{h}{\eta_{i}}=1$: $x$ has a probability $\eta_{i}$ to follow an exponential distribution with rate $\pi_{i}$. For an exponential distribution with rate $\pi_{i}$, its mean is $\frac{1}{\pi_{i}}$. The mean of $x$ is
\begin{align}\label{equa-hyper-mean}
s=\sum\nolimits_{i=1}^{h}{\frac{\eta_{i}}{\pi_{i}}}=1
\end{align}
and the second moment of $x$ is
\begin{align}\label{equa-hyper-squared-mean}
E[x^{2}] = \sum\nolimits_{i=1}^{h}{\frac{2}{\pi_{i}^{2}}\cdot \eta_{i}}.
\end{align}

\subsection{Optimal SLA Delays}

The actual experienced job delays of the $L$ SLAs are $t_{1}, \cdots, t_{L}$. As described in (\ref{qos-constraint}), $t_{l}$ should be no larger than the SLA delay $\varphi_{l}$. The delay of the first SLA is $T$. Intuitively, we should keep the other SLA delays as small as possible, i.e., $\varphi_{l}=t_{l}$ for all $l\in [2, L]$, in order to maximize the revenue. In fact, by doing so, we can make every SLA price as high as possible, and we now rigorously prove this by analyzing the structure of the SLA prices in Definition~\ref{def-optip}.

\begin{proposition}\label{proposi-optimal-delays}
In order to maximize the revenue, we have $\varphi_{l}=t_{l}$ for all $l\in [2, L]$.
\end{proposition}

\subsection{Performance Bounds}

In this subsection, we will study the performance of the proposed service system respectively built on the PBS and SMS architectures. Recall that $G$ denotes the revenue of the service system of this paper and we denote by $G_{od}$ the revenue of an on-demand service system. The viability of our service system can be mainly indicated by the ratio of $G$ to $G_{od}$, denoted by $\kappa$; $\kappa-1$ represents how much the revenue $G_{od}$ is improved by when our service system is used. It is difficult to give a closed form of the optimal $G$ since this involves solving a system of non-linear equations. We thus seek to give a bound of $\kappa$.

For the PBS-based service system, we will get an upper bound of $\kappa$ that is close to one. This implies that, at best, it can marginally outperform the on-demand service system, which will discourage the adoption of a PBS-based service system. For the SMS-based service system, we will get a lower bound of $\kappa$ that is significantly larger than one. This implies that the SMS-based service system can significantly outperform the on-demand service system, which will support the use of a SMS-based service system by CSPs. Finally, we will give an optimal algorithm to maximize the revenue of a SMS-based service system.

\subsubsection{\textbf{A Performance Bound of the PBS-based Service System}}
\label{Sec.optimal-delay-PBS}

When a PBS-based service system is considered, we denote by $G_{pbs}$ its revenue. We will derive an upper bound of the ratio of $G_{pbs}$ to $G_{od}$. For the standard on-demand service model, it has a fixed price $p$ and guarantees a small delay of at most $T$. A CSP's revenue is maximized when the delay of the first SLA is $T$ and we have by (\ref{equa-waiting-time-on-demand}) that the corresponding job arrival rate at a single server is as follows:
\begin{align}\label{performance-od}
\lambda_{od} = \frac{T}{A+T},
\end{align}
where $A=0.5\cdot E[x^{2}]$. Further, the maximum revenue that an on-demand service model can achieve is
\begin{align}\label{equa-revenue-on-demand}
G_{od} = m\cdot p\cdot\lambda_{od}\cdot s = m\cdot p\cdot\frac{T}{A+T}.
\end{align}

For the PBS-based service system, we have the following analysis. All jobs of different SLAs are executed on the $m$ servers. The first SLA offers service at a fixed price $p$ and guarantees a small delay of at most $T$, and we have by (\ref{equa-waiting-time-random-0}) that
$\varphi_{1} = \lambda\cdot A /\left(1-\hat{\lambda}_{1}\right) \leq T$, where $0<\hat{\lambda}_{1}<\lambda<1$. Thus, we get $\lambda<T/A$. A CSP's revenue is given in (\ref{equa-objective-fun}) and we can get an upper bound of $G_{pbs}$:
\begin{align}\label{performance-upper-bound-pbs}
G_{pbs} = \sum\limits_{l=1}^{L}{p_{l}\cdot m\cdot\lambda_{l}}\leq p\cdot m\cdot\sum\limits_{l=1}^{L}{\lambda_{l}}=p\cdot m\cdot \lambda<p\cdot m\cdot \frac{T}{A},
\end{align}
where $p_{l}\leq p$ for all $l\in [1, L]$. It follows from (\ref{performance-od}) and (\ref{performance-upper-bound-pbs}) that

\begin{proposition}\label{proposi-performance-upper-bound-pbs}
The performance of a PBS-based service model is upperly bounded by $1+\frac{T}{A}$ times the optimal performance of the standard on-demand service model, in terms of the revenue, where $A=0.5\cdot E[x^{2}]$.
\end{proposition}

\begin{figure}[t]
  \centering
  \includegraphics[width=3.3in]{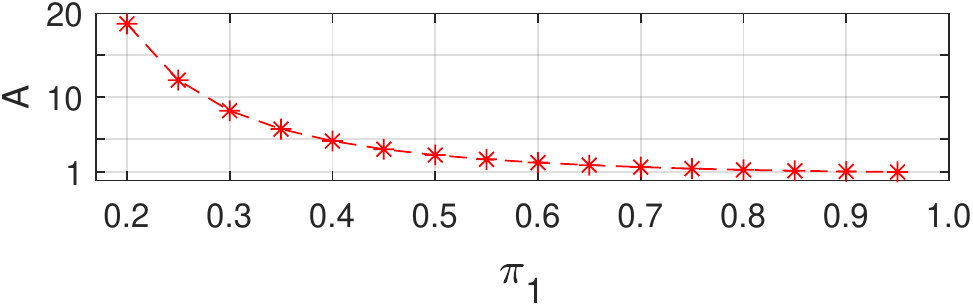}
  \caption{\textbf{The Value of $A$ under Varying $\pi_{1}$.}}\label{Fig-upper-bound}
\end{figure}

When $x$ follows an exponential distribution, we have $A=1$ by (\ref{equa-exp}). When $x$ follows a hyperexponential distribution, we use an example in \cite{Mukherjee} to set $h=2$, $\eta_{1}=0.75$ $\eta_{2}=0.25$; we let $\pi_{1}\in (0, 1)$, which represents more jobs have relatively smaller service times. We vary the value of $\pi_{1}$ from 0.2 to 0.95 with a step size 0.05, and compute the corresponding value of $\pi_{2}$ by (\ref{equa-hyper-mean}); then, we can get the value of $A$ by (\ref{equa-hyper-squared-mean}), which is illustrated by the red stars in Fig.~\ref{Fig-upper-bound}, where $A>1$. In both cases, we can conclude by Proposition~\ref{proposi-performance-upper-bound-pbs} that, the upper bound in Proposition~\ref{proposi-performance-upper-bound-pbs} is at most $1+T$, and the PBS-based service system can only outperform the standard on-demand service system marginally, since the delay of the first SLA $T$ is small.

\subsubsection{\textbf{A Performance Bound of the SMS-based Service System}}
\label{sec.performance-bound-sms}

%\vspace{0.12em}\noindent{\color{blue}Analytical results: continuous distribution}

When a SMS-based service system is considered, we denote by $G_{sms}$ its revenue, and by $G_{sms}^{*}$ its optimal revenue where $G_{sms}^{*}\geq G_{sms}$. In cloud markets, the total number of servers is large so that the revenue from a single server could be negligible, in comparison with the total revenue. Thus, to give a closed form of the lower bound, we relax in this subsubsection the constraint that the number of servers assigned to each SLA is integer and allow the number to be fractional; the total revenue after relaxation approximates the total revenue of an integer solution. Further, we have $m_{i}\cdot\lambda_{i}=\Lambda_{i}$, and $\lambda_{i}$ and $\varphi_{i}$ satisfy the relation (\ref{equa-waiting-time-separated-random}); for $i\in [1, L]$, the number of servers assigned to the $i$-th SLA is as follows:
\begin{align}\label{equa-number-servers-two-slas}
m_{i} = \frac{\Lambda_{i}\cdot (\varphi_{i}+A)}{\varphi_{i}},
\end{align}
where $A=0.5\cdot E[x^{2}]$. Furthermore, it is known that there are many applications whose workload is delay-tolerant, as illustrated by the prosperity of spot market \cite{amazon-spot-users}; a CSP like Amazon EC2 or Microsot Azure also has the ability to adjust the provision of servers to properly satisfy the needs of users.

We consider a specific setting of the service system where two SLAs are offered respectively for latency-critical and delay-tolerant jobs; the corresponding revenue can be viewed as a lower bound of $G_{sms}^{*}$. The setting is as follows: (\rmnum{1}) we choose some $\alpha^{\prime}\in \left(\underline{\alpha},\, \overline{\alpha}\right)$ such that all customers with $\alpha$ larger than $\alpha^{\prime}$ will be processed under the first SLA and the others are processed under the second SLA (i.e., $\hat{\alpha}_{2}=\alpha^{\prime}$), and (\rmnum{2}) the CSP intends to adapt its capacity (i.e., the value of $m_{2}$) to guarantee that the delay $\varphi_{2}$ of the second SLA is set to some value $\varphi_{2}^{\prime}$; the $\alpha^{\prime}$ and $\varphi_{2}^{\prime}$ are system parameters set by the CSP. $\alpha^{\prime}$ determines the proportion of the arriving jobs to be processed under each SLA. Let $\Phi_{1}=\Phi\cap (\alpha^{\prime}, \overline{\alpha}]$, and $\Phi_{2}=\Phi-\Phi_{1}$; we have that the job arrival rates for the first and second SLAs are respectively: $\Lambda_{1} = \sum\nolimits_{\alpha\in\Phi_{1}}{P(\alpha)}\cdot \Lambda  \,\text{ and }\, \Lambda_{2}=\Lambda-\Lambda_{1}$.
By Proposition~\ref{theo-sequence-1}, the prices of the first and second SLAs are $p_{1}=p$ and $p_{2}=p_{1}+(u(\hat{\alpha}_{2},\varphi_{2})-u(\hat{\alpha}_{2},\varphi_{1}))=u(\hat{\alpha}_{2}, \varphi_{2})$ where $\varphi_{1}=T$. We have that the CSP's total revenue is
\begin{align}\label{equa-revenue-sms-two-slas}
G_{sms} = p\cdot \Lambda_{1} + u\left(\alpha^{\prime},\,\varphi_{2}^{\prime}\right)\cdot \Lambda_{2}.
\end{align}
If the CSP only provides on-demand service, the (optimal) revenue $G_{od}$ is as follows:
\begin{align}\label{equa-revenue-on-demand-two-SLAs}
G_{od} = m\cdot p\cdot \lambda_{od} = \left( m_{1} + m_{2} \right)\cdot p\cdot \frac{T}{A+T},
\end{align}
where $\lambda_{od}$ is given in (\ref{performance-od}). The below conclusion follows from (\ref{equa-revenue-sms-two-slas}) and (\ref{equa-revenue-on-demand-two-SLAs}):

\begin{proposition}\label{proposi-lower-bound}
The optimal revenue $G_{sms}^{*}$ of a SMS-based service system is at least $\kappa^{\prime}$ times the revenue of an on-demand service system where
\begin{align}\label{equa-closed-form-lower-bound}
\kappa^{\prime}=\frac{G_{sms}}{G_{od}} \geq  \frac{\left(p\cdot \Lambda_{1} + u\left(\alpha^{\prime},\,\varphi_{2}^{\prime}\right)\cdot \Lambda_{2}\right)\cdot (A+T)}{\left( m_{1}+m_{2} \right)\cdot p\cdot T},
\end{align}
where $m_{1}$ and $m_{2}$ are given in (\ref{equa-number-servers-two-slas}), $A=0.5\cdot E[x^{2}]$, and $\alpha^{\prime}$ and $\varphi_{2}^{\prime}$ are system parameters set by the CSP.
\end{proposition}

\begin{figure}[t]
  \centering
  \includegraphics[width=3.1in]{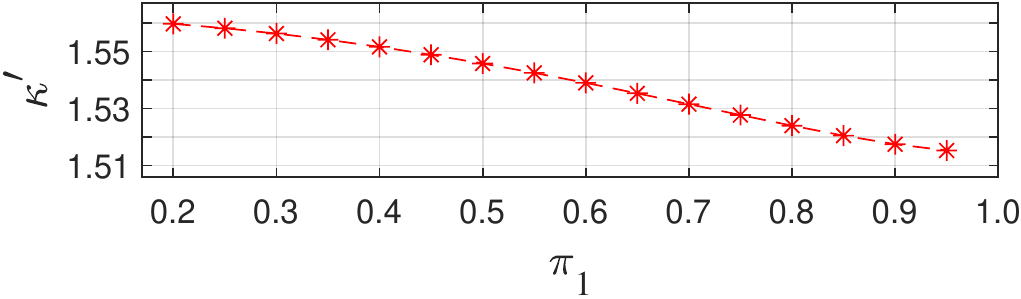}
  \caption{\textbf{The Lower Bound $\kappa^{\prime}$ under Varying $\pi_{1}$.}}\label{Fig-lower-bound}
\end{figure}

Proposition~\ref{proposi-lower-bound} provides a closed form of the lower bound $\kappa^{\prime}$ of the ratio of $G_{sms}^{*}$ to $G_{od}$, and $\kappa^{\prime}-1$ represents the minimum revenue improvement brought by a SMS-based service system. One of its advantages lies in that one can get the value of $\kappa^{\prime}-1$ through an easy computation of (\ref{equa-closed-form-lower-bound}), without the need of executing a procedure that one may need to take some effort to implement. Now, we give an instance of Proposition~\ref{proposi-lower-bound}. We use the WTP function in (\ref{equa-wtp}) where $\beta=3$:
\begin{align}\label{equa-specific-WTP-fun}
u(\alpha, \varphi)=p\cdot\left(1-(\alpha\cdot(\varphi-T))^{3}\right), \enskip t\in [T, +\infty).
\end{align}
The parameter $\alpha^{\prime}$ is set to be such that a significant portion of jobs (e.g., half jobs) are processed under each SLA. Since there are many delay-tolerant jobs in cloud markets, $\hat{\varphi}_{0}=\frac{1}{\alpha^{\prime}}+T$ can be much larger than $T$ and it is the minimum delay under which the WTPS of the customers of the second SLA will become zero. Correspondingly, the delay $\varphi_{2}$ of the second SLA is set to $\frac{\hat{\varphi}_{0}+T}{2}$; this leads to that the price $p_{2}$ of the second SLA will be $0.875\cdot p$, which is not far from the on-demand price $p$. We set $\Lambda_{1}=\Lambda_{2}=0.5\cdot\Lambda$. In this case, we have that
\begin{align}\label{equa-lower-bound-specific}
\kappa^{\prime}=\frac{G_{sms}^{*}}{G_{od}} \geq  \frac{1.875\cdot \left(1+\frac{A}{T}\right)}{\left(2+\frac{A}{T}+2\cdot \frac{A}{\hat{\varphi}_{0}}\right)}.
\end{align}
The lower bound of (\ref{equa-lower-bound-specific}) decreases in $T$ and increases in $\hat{\varphi}_{0}$; we set $T$ to a larger value 0.05 and $\hat{\varphi}_{0}$ to 0.5. In this case, when the service time $x$ follows an exponential distribution, we have $A=1$ by (\ref{equa-exp}) and $\kappa^{\prime}\geq 1.514$. When $x$ follows a hyperexponential distribution, we still use the setting in Section~\ref{Sec.optimal-delay-PBS} and the value of $\kappa^{\prime}$ is illustrated in Fig.~\ref{Fig-lower-bound} where $\kappa^{\prime}\geq 1.515$. For both distributions, we have that, if the CSP adopts the proposed service model under the SMS architecture, the revenue can be at least 1.5 times the optimal revenue of the pure on-demand service system, with a remarkable improvement.

\subsection{Optimally Configuring the SMS-based Service System}
\label{sec.optimal-delay-SMS}

\begin{algorithm}[t]
%\DontPrintSemicolon
\SetKwInOut{Begin}{Begin}
\SetKwInOut{Input}{Input}
%\footnotesize{}
    \setlength\itemsep{0.5em}

    $G^{*}\leftarrow 0$, $\mathcal{A}^{\prime}\leftarrow \mathcal{A}$, $\mathcal{M}^{\prime}\leftarrow \mathcal{M}$\tcp*{\footnotesize{$G^{*}$: record the current optimal revenue; $\mathcal{A}^{\prime}$ and $\mathcal{M}^{\prime}$: record the tuples unconsidered respectively in $\mathcal{A}$ and $\mathcal{M}$}}

\While{$\mathcal{M}^{\prime}\neq \emptyset$}{

    Get a tuple $(i_{1}, i_{2}, \cdots, i_{L+1})$ from $\mathcal{M}^{\prime}$, and the $l$-th module is assigned $m_{l}=i_{l+1}-i_{l}$ servers\;

    \While{$\mathcal{A}^{\prime}\neq \emptyset$}{

        Get a tuple $seq=(\alpha_{1}, \alpha_{2}, \cdots, \alpha_{L+1})$ from $\mathcal{A}^{\prime}$\;

        Compute the job arrival rate $\Lambda_{l}$ of the $l$-th SLA by Equation (\ref{equa-arrival-rate-l}) and Proposition~\ref{theo-market-segmentation}\;

        For all $l\in [1, L]$, compute the actual job delay $t_{l}$ of the $l$-th SLA using (\ref{equa-waiting-time-separated-random})\;

        \If{$\varphi_{1}\leq T < \varphi_{1}< \varphi_{2}<\cdots<\varphi_{L}$}{
            \tcp{\footnotesize{The delay of the first SLA is no larger than $T$ and the SLA delays are increasing}}

            Set the delay $\varphi_{l}$ of the $l$-th SLA to $t_{l}$ for all $l\in [2, L]$, and $\varphi_{1}$ to $T$\;

            Use Proposition~\ref{theo-sequence-1} to compute the optimal prices of SLAs $p_{1}, p_{2}, \cdots, p_{L}$\;

            Compute the revenue $G$ by (\ref{equa-objective-fun}), where $w_{l}=\Lambda_{l}\cdot s=m\cdot\lambda_{l}\cdot s$\;

            \If{$G>G^{*}$}{

                $G^{*}\leftarrow G$,\, $\varphi_{l}^{*}\leftarrow \varphi_{l}$, $p_{l}^{*}\leftarrow p_{l}$, $m_{l}\leftarrow m_{l}^{*}$, for all $l\in [1, L]$\tcp*{\footnotesize{record the optimal SLA delays and prices, and division of servers}}
            }

         }

        Delete $seq$ from $\mathcal{A}^{\prime}$\;
    }
    Delete the tuple $(i_{1}, i_{2}, \cdots, i_{L+1})$ from $\mathcal{M}^{\prime}$\;
}

\caption{Optimal Parameter Configuration}\label{Algo-OptiDelay-separated}
\end{algorithm}

In this subsection, we will give a procedure to determine the optimal SLA delays and prices of a SMS-based service system, in order to maximize the revenue. The delays and prices are determined by the market segmentation $\hat{\alpha}_{1}, \hat{\alpha}_{2}, \cdots, \hat{\alpha}_{L+1}$, and the numbers of servers assigned to different SLAs $m_{1}, m_{2}, \cdots, m_{L}$. Specifically, as shown in Proposition~\ref{theo-SLA-assignment-maximum}, the sequence $\hat{\alpha}_{1}, \hat{\alpha}_{2}, \cdots, \hat{\alpha}_{L+1}$ determines the job arrival rate of each SLA by (\ref{equa-arrival-rate-l}). The numbers $m_{1}, m_{2}, \cdots, m_{L}$ determine the delays of SLAs $\varphi_{1}, \varphi_{2}, \cdots, \varphi_{L}$ by (\ref{equa-waiting-time-separated-random}), which further determine the prices of SLAs by Proposition~\ref{theo-sequence-1}. Thus, our decision variables are $\hat{\alpha}_{2}, \cdots, \hat{\alpha}_{L}$ and $m_{1}, \cdots, m_{L}$ with the aim of maximizing the revenue, where $\hat{\alpha}_{1}=\overline{\alpha}$, $\hat{\alpha}_{L+1}=\underline{\alpha}$, and $\sum_{l=1}^{L}{m_{l}}=m$.

Now, we give a procedure to determine the optimal decision variables under the SMS architecture. $\hat{\alpha}_{1}, \hat{\alpha}_{2}, \cdots, \hat{\alpha}_{L+1}$ uniquely corresponds to an element in the following set
\begin{equation*}
\begin{split}
\mathcal{A} = \{ (\alpha_{1}, \alpha_{2}, \cdots, \alpha_{L+1}) \,|\, \overline{\alpha} = \alpha_{1} > \alpha_{2} > & \cdots > \alpha_{L+1} = \underline{\alpha}, \\
& \alpha_{2}, \alpha_{3}, \cdots, \alpha_{L}\in \Phi \},
\end{split}
\end{equation*}
where $\hat{\alpha}_{l}=\alpha_{l}$ for all $l\in [1, L+1]$. $m_{1}, m_{2}, \cdots, m_{L}$ uniquely correspond to an element in the following set
\begin{align*}
\mathcal{M} = \left\{(i_{1}, i_{2}, \cdots, i_{L+1}) \,|\, 0 = i_{1} < i_{2}< \cdots < i_{L+1}=m \right\}.
\end{align*}
The number $m_{l}$ is set to $i_{l+1}-i_{l}$ for all $l\in [1, L]$. We can give a procedure, presented as Algorithm~\ref{Algo-OptiDelay-separated}, to determine the optimal tuples in $\mathcal{A}$ and $\mathcal{M}$ such that the CSP achieves the maximum revenue; then, the corresponding delays and prices under these two tuples will be the optimal ones, and we have the following conclusion.

\begin{proposition}\label{proposi-opti-configuration}
Algorithm~\ref{Algo-OptiDelay-separated} gives the optimal delays and prices of SLAs, and its time complexity is $\mathcal{O}\left(m^{L-1}\cdot n^{L-1}\right)$.
\end{proposition}

\begin{figure*}[t]
  \centering
  \includegraphics[width=2.55in]{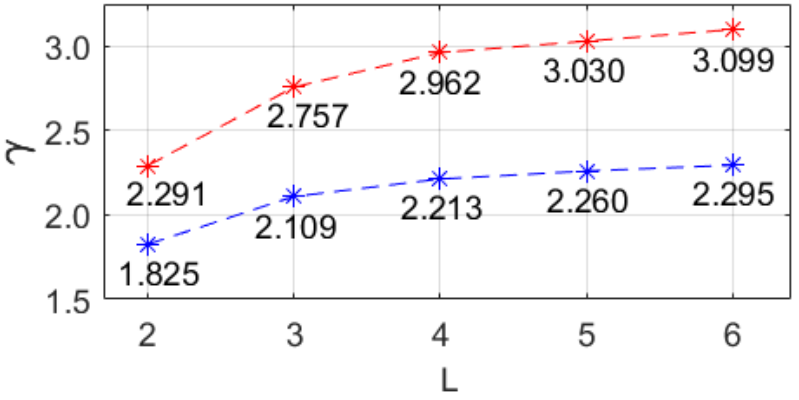}%\\
  \hspace{0.15cm}\includegraphics[width=2.55in]{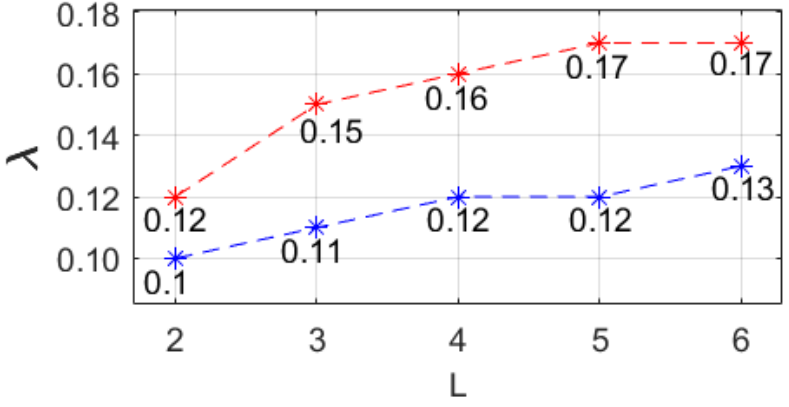}%\\
  \caption{\textbf{Revenue Improvement}: the red (resp. blue) stars are for the case of low (resp. high) delay-tolerance; the left subfigure illustrates the maximum revenue improvement under a given number of SLAs $L$, while the right subfigure illustrates the corresponding average load per server; in the on-demand service system, the average load per server is 0.0476.}\label{Fig-optimal-revenue-improvement}
\end{figure*}

\section{Numerical Results}
\label{sec.performance-evaluation}

In this section, we numerically show the revenue improvement that a SMS-based service system achieves over the standard on-demand service system. {Besides, we adapt the architecture of \cite{Abhishek12,Dierks19} to the service model of this paper and compare it with the SMS-based service system; the related results and analysis are put in the Appendix.}

%Our analysis in Section~\ref{sec.serving-architecture} discourages the adoption of a PBS-based service system and supports the use of a SMS-based service system, whose performance will be further evaluated in this section. We will use Algorithm~\ref{Algo-OptiDelay-separated} to give some numerical results under a wide range of market conditions. This helps understand the performance of a SMS-based service system, in comparison with the standard on-demand service system. {\color{blue}Furthermore, we also adapt the architecture of \cite{Abhishek12,Dierks19} to the service model of this paper and give some numerical results to take some comparison with the SMS-based service system.}

%In addition, we also compar

\subsection{Experimental Setting}

There are a total of $m$ servers and the WTP function is given in (\ref{equa-specific-WTP-fun}). The on-demand price $p$ (i.e., the price $p_{1}$ of the first SLA) is normalized as 1, and its delay $T$ is 0.05. Given a delay-cost type $\alpha$, let $\varphi_{0}^{\prime}=\frac{1}{\alpha}$ and a customer's WTP becomes zero when the delay $\varphi_{0}=\varphi_{0}^{\prime}+T$, and each $\alpha$ uniquely corresponds to a $\varphi_{0}$. There are $n=50$ types of customers and for all $i\in [1, n]$ the WTP of the $i$-th type of customers becomes zero when the delay is $\varphi_{0,i}=T+\varphi_{0,i}^{\prime}$; here, $\varphi_{0,i}^{\prime}=\epsilon$ if $i=1$ and $\varphi_{0,i}^{\prime}=(i-1)\cdot \delta$ otherwise, where $\epsilon$ is arbitrarily small. We have $\varphi_{0,1}<\varphi_{0,2}<\cdots<\varphi_{0,50}$. The first type of customers is the most delay-sensitive and its WTP becomes zero even if the delay is slightly larger than $T$. The value of $\delta$ determines the delay-tolerance of the population, and if it is large, the population has a high delay-tolerance. We consider two cases where the delay-tolerance is low and high respectively: (\rmnum{1}) $\delta=0.02$ and (\rmnum{2}) $\delta=0.04$.

The mean arrival rate of the jobs of all types is $\Lambda$; the service time of jobs follows an exponential distribution and their mean is normalized as one, i.e., $s=1$. Customers are independently and uniformly distributed over the $n$ types, and the mean job arrival rate of each type is $\frac{\Lambda}{n}$. Then, $\rho=\frac{\Lambda}{m}\cdot s=\lambda$ denotes the average load per server when all $m$ servers are considered.
%$\Lambda_{l}$ denotes the job arrival rate of the $l$-th SLA and is given in (\ref{equa-arrival-rate-l}); $\rho_{l}=\frac{\Lambda_{l}}{m_{l}}\cdot s=\lambda_{l}$ denotes the average load per server of the $l$-th SLA.
We denote by $G_{sms}^{*}$ the optimal revenue achieved by Algorithm~\ref{Algo-OptiDelay-separated}. In an on-demand service system, $G_{od}$ denotes its revenue and is defined in (\ref{equa-revenue-on-demand}); $\lambda_{od}$ denotes the maximum load per server and it equals 0.0476 since $t\leq T$ in (\ref{equa-waiting-time-on-demand}); hence, the maximum revenue that a CSP can obtain from a single server is also 0.0476. {The following ratio is the main performance metric in our experiments:
$$
\gamma = G_{sms}^{*}/G_{od}.
$$
Specifically, if $\gamma>1$, the SMS-based service system will outperform the on-demand system; the larger the value of $\gamma$, the higher the revenue improvement.}

%The revenue improvement brought by a SMS-based service system can be indicated by the ratio below:
%$$
%\gamma = G_{sms}^{*}/G_{od},
%$$
%which is a main performance metric in our experiments.

\subsection{Numerical Results}

The service model of this paper can be viewed as a complement to the on-demand service, and it can attract potential delay-tolerant customers from the market and improve the revenue efficiency, i.e., the average revenue per server.
%Although we focus in this paper on the revenue improvement of the proposed model to the on-demand service model, it also brings further opportunity for a CSP to lower its on-demand price $p$, which can even attract more latency-critical customers from its competitors, as well as those delay-tolerant customers.
In practice, a CSP like Amazon EC2 or Microsoft Azure often has rich capital and can adapt its capacity to accept and serve all arriving jobs and maintain its load per server at a desired level. %it may more value its reputation and profitability.

\begin{figure*}[t]
  \centering
  \includegraphics[width=2.5in]{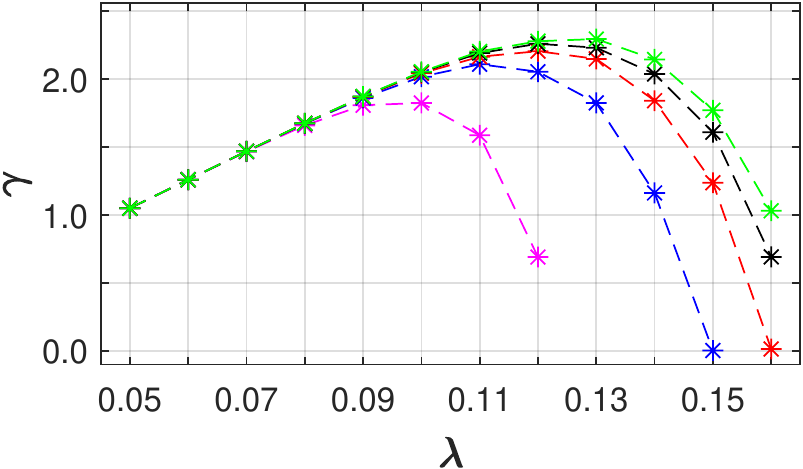}%\\
  \hspace{0.25cm}\includegraphics[width=2.5in]{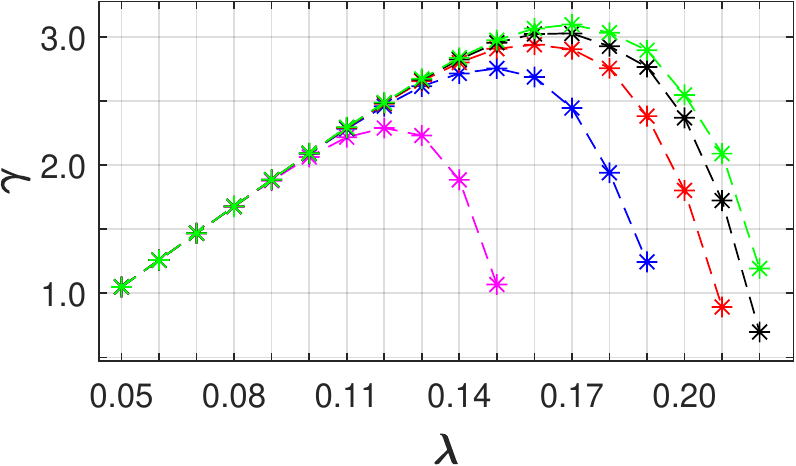}%\\
  \caption{\textbf{Revenue Improvement $\gamma$ under Varying Load $\lambda$}: (\rmnum{1}) the left and right subfigures correspond to the low and high delay-tolerance cases respectively; (\rmnum{2}) the magenta, blue, red, black and green stars denote the revenue improvement $\gamma$ in the case of two, three, four, five and six SLAs respectively.}\label{Fig-revenue-improvement-varying-rate}
\end{figure*}

%\begin{table*}[t]
%	\centering
	%	\caption{\textbf{The optimal revenue improvement $\gamma$ and the corresponding average load per server $\lambda$} when the number of SLAs is $L$: the first and second cases correspond to the cases of low and high delay-tolerance respectively; in the on-demand service system, the load per server is 0.0476.}
%		\label{revenue_improvement_1}
%	\begin{threeparttable}[b]
%		\begin{tabular}{|C{0.9cm}|C{0.9cm}|C{0.9cm}|C{0.9cm}|C{0.9cm}|C{0.9cm}|C{0.9cm}|C{0.9cm}|C{0.9cm}|C{0.9cm}|C{0.9cm}|}	
%			\hline
%\cline{1-7} \multicolumn{1}{ |c| }{} &  \multicolumn{5}{ c| }{Case 1} & \multicolumn{5}{ c| }{Case 2}\\ \hline
%     $L$     &    $2$   &  $3$      &  $4$       &  5     &  6     &  $2$       &     $3$      &     $4$      &     5      &   6    \\ \hline			
%   $\rho$    &   0.1    &   0.11    &    0.12    &  0.12  &  0.13  &    0.12    &    0.15      &    0.16      &     0.17   &   0.17     \\ \hline
%   $\gamma$  & 1.825    &   2.109   &    2.213   &  2.260 &  2.295 &   2.291    &    2.757     &    2.962     &    3.030   &   3.099 \\ \hline
%		\end{tabular}
%	\end{threeparttable}
%\end{table*}

\begin{figure*}[t]
  \centering
  \includegraphics[width=2.5in]{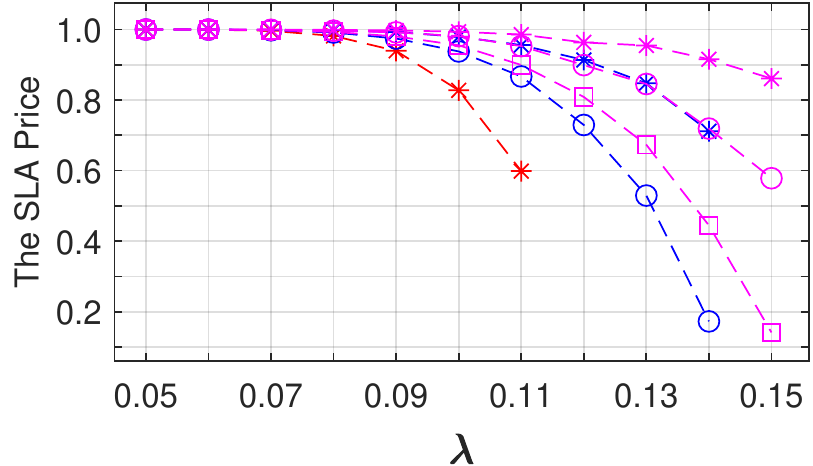}%\\
  \hspace{0.25cm}\includegraphics[width=2.5in]{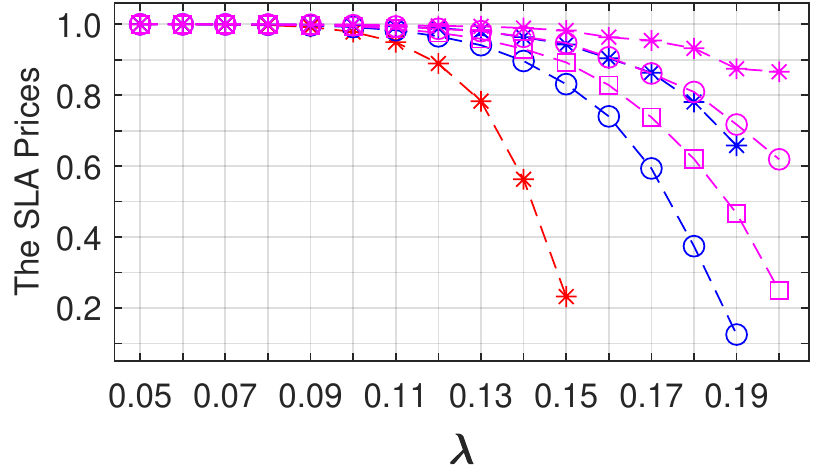}%\\
  \caption{\textbf{The SLA Prices under Varying Load $\lambda$}: (\rmnum{1}) the left and right subfigures correspond to the low and high delay-tolerance cases respectively; (\rmnum{2}) in each subfigure, the red, blue and magenta markers denote the results when $L=2,\,3,\,4$ respectively; (\rmnum{3}) the markers "stars", "circles" and "squares" denote the SLA prices of the second, third and fourth SLAs respectively; the price of the first SLA is one.}\label{Fig-SLA-prices}
\end{figure*}

\begin{figure*}[t]
  \centering
  \includegraphics[width=2.5in]{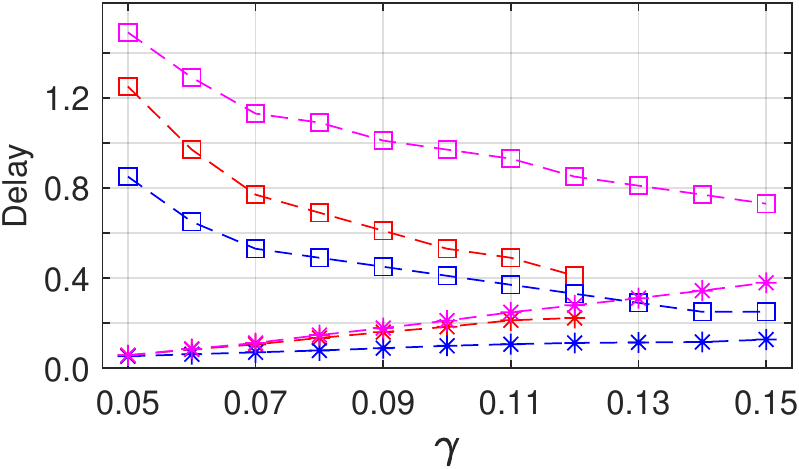}%\\
  \hspace{0.25em}
  \includegraphics[width=2.5in]{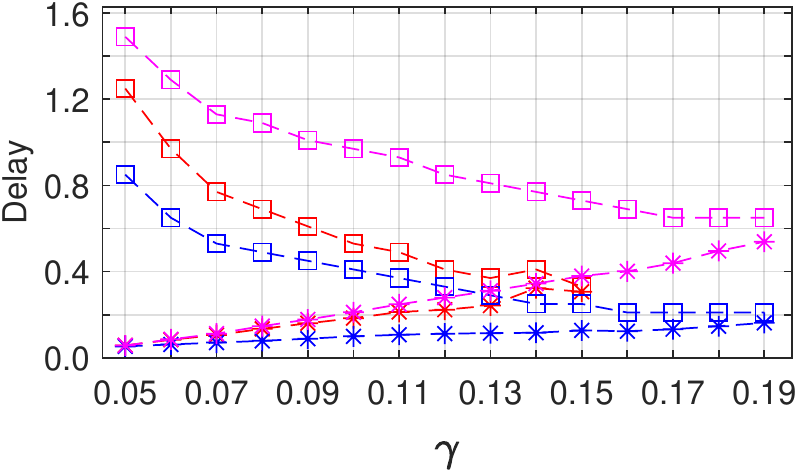}%\\
  \caption{\textbf{The SLA Delays under Varying Load $\lambda$}: (\rmnum{1}) the left and right subfigures correspond to the low and high delay-tolerance cases respectively; (\rmnum{2}) the stars illustrate the SLA delays $\varphi_{l}$ while the squares illustrate the value of $\varphi_{0,i_{l}}$; (\rmnum{3}) the red markers are for the second SLA when offering two SLAs; (\rmnum{4}) the blue and magenta markers are respectively for the second and third SLAs when offering three SLAs.}\label{Fig-SLA-delay}
\end{figure*}

%\begin{figure*}[t]
%  \centering
%  \includegraphics[width=2.5in]{fig-number-servers-23-low.pdf}%\\
%  \hspace{0.25cm}\includegraphics[width=2.5in]{fig-number-servers-4-low.pdf}%\\
%  \caption{\textbf{Server Assignment in the Low Delay-Tolerance Case}: in both subfigures, the markers "stars", "circles" and "squares" are respectively for the case of two, three and four SLAs; the red, blue, magenta, and black markers denote the number of servers assigned to the second, third and fourth SLAs, respectively.}\label{Fig-number-server-low}
%\end{figure*}

%\vspace{0.25em}\noindent\textbf{Revenue Improvement.}
\subsubsection{\textbf{Revenue Improvement}}

In Section~\ref{sec.performance-bound-sms}, we have given a lower bound of the performance in (\ref{equa-lower-bound-specific}), and consider the setting that two SLAs are offered and each SLA is assigned half the jobs. When it is further concretized by our experimental setting, we have $\hat{\varphi}_{0}=0.05+25\cdot \delta$. In the low delay-tolerant case, $\delta=0.02$ and $\hat{\varphi}_{0}=0.55$; the revenue improvement $\gamma$ is 1.536. In the high delay-tolerant case, $\delta=0.04$ and $\hat{\varphi}_{0}=1.05$; $\gamma$ is 1.647.

In the rest of this section, we fix the number of servers $m=100$ and allocate a proper proportion of servers to each SLA. We vary the average load per server $\lambda$ that increases from 0.05 with a step size 0.01, and calculate the revenue improvement $\gamma$. The value of $\gamma$ varies under different load $\lambda$. The maximum revenue improvement under a given number of SLAs $L$ is summarized in Fig.~\ref{Fig-optimal-revenue-improvement} (left), ranging from 182.5\% to 309.9\%; the corresponding optimal $\lambda$ is given in Fig.~\ref{Fig-optimal-revenue-improvement} (right). From the figure, we can see that (\rmnum{1}) the larger the number $L$ of SLAs, the higher the revenue improvement $\gamma$, and (\rmnum{2}) the higher the delay-tolerance, the higher the revenue improvement. In the low delay-tolerance case, when the number of SLAs offered by a CSP varies from two to six, the revenue improvement increases from 182.5\% to 226.0\%. The revenue improvement is remarkable even when $L=2$. In the high delay-tolerance case, the revenue improvement is 229.1\% even when $L=2$. In both low and high delay-tolerance cases, when $L\geq 4$, the revenue improvement increases only marginally as $L$ increases. This may imply that, in practice, offering two or three SLAs may be enough.

%\begin{figure*}[t]%
%  \centering
%  \includegraphics[width=2.5in]{fig-number-servers-23-high.pdf}%\\
%  \hspace{0.25cm}\includegraphics[width=2.5in]{fig-number-servers-4-high.pdf}%\\
%  \caption{\textbf{Server Assignment in the High Delay-Tolerance Case}: the meanings of the markers and colors are the same as Figure~\ref{Fig-number-server-low}.}\label{Fig-number-server-high}
%\end{figure*}

%\vspace{0.25em}\noindent\textbf{Further Observation.}
\subsubsection{\textbf{Further Observation}}

In the following, we illustrate some detailed numerical results to help us understand the features of an optimal parameter configuration.

First, we describe the general features. By Proposition~\ref{theo-sequence-1}, there exists a sequence $1 = i_{1} < i_{2} < \cdots <$ $i_{L}$ $\leq n$ such that the $l$-th SLA is assigned the customers whose $\varphi_{0,i}$ is such that $i\in [i_{l},\, i_{l+1})$ if $l\in [1, L-1]$ and $i\in [i_{L},\, n]$ if $l=L$; here, we have $\hat{\alpha}_{l}=$ $1/\varphi_{0, i_{l}}^{\prime}$. For all $l\in [2, L]$, the price $p_{l}$ of the $l$-th SLA equals $p_{l-1}$ minus the difference $u(\hat{\alpha}_{l}, \varphi_{l-1})-u(\hat{\alpha}_{l}, \varphi_{l})$ where $\varphi_{l-1}<\varphi_{l}$. Roughly, the revenue is the average price times the load of the $m$ servers. To maximize the revenue, we need keep the SLA prices high, and the sequence $i_{2}, i_{3}, \cdots, i_{L}$ should be selected in a way such that, for all $l\in [2, L]$,
\begin{enumerate}
   \item [(\rmnum{1})] the SLA delay $\varphi_{l}$ is significantly smaller than $\varphi_{0, i_{l}}$;
   \item [(\rmnum{2})] the difference of $\varphi_{l}$ and $\varphi_{l-1}$ is small;
   \item [(\rmnum{3})] the value of $\varphi_{0, i_{l}}$ is as large as possible;
   \item [(\rmnum{4})] the SLA delay $\varphi_{l}$ is significantly larger than $T$.
\end{enumerate}
When the delay is small, the WTP decreases slowly, as explained in Section~\ref{sec.delay-cost-curve}. The first two points guarantee that $p_{l}$ is not far from the on-demand price $p$. By (\ref{equa-waiting-time-separated-random}), the last two guarantee that, the load $\lambda_{l}$ per server of the $l$-th SLA is significantly larger than $\lambda_{od}$, leading to a larger overall load $\lambda$ per server.

Second, the above features are also embodied in our numerical results. The revenue improvement under varying load are illustrated in Fig.~\ref{Fig-revenue-improvement-varying-rate}. The corresponding SLA prices and delays are given in Fig.~\ref{Fig-SLA-prices} and \ref{Fig-SLA-delay}. Given the number of SLAs $L$, the revenue improvement $\gamma$ always increases until the load $\lambda$ increases to some threshold; afterwards, $\gamma$ begins to decrease since every server has a too heavy load. As illustrated in Fig~\ref{Fig-SLA-delay}, if $\lambda$ is too large, the SLA delay $\varphi_{l}$ will be large and close to $\varphi_{0,i_{l}}$; then, the WTPs of customers are low, as well as the SLA price, as illustrated in Fig.~\ref{Fig-SLA-prices}; thus, $\gamma$ becomes smaller even if more workload is processed.

For example, in the low delay-tolerance case with $L=2$, the optimal $\gamma$ is achieved when the load $\lambda$ is 0.1, as shown in Fig.~\ref{Fig-optimal-revenue-improvement} (right). As the load $\lambda$ increases from 0.05 to 0.1, $\gamma$ keeps increasing, as illustrated by the magenta curve in Fig.~\ref{Fig-revenue-improvement-varying-rate} (left); afterwards, $\gamma$ begins to decrease. As illustrated by the red curve in Fig.~\ref{Fig-SLA-delay} (left), when the load $\lambda$ is 0.12, the SLA delay $\varphi_{2}$ is 0.2228, which is close to $\varphi_{0,i_{2}}=0.23$; then, the SLA price $p_{2}$ is 0.1149. In contrast, when $\lambda=0.1$, the SLA delay $\varphi_{2}=0.1836$, which is significantly smaller than $\varphi_{0,i_{2}}=0.29$. Thus, to maintain a large $\gamma$, the average load $\lambda$ per server should be maintained at a proper level by adjusting the total number of servers $m$.

\section{Conclusion}
\label{sec.conclusion}

%%%%%%%%%%%%%%%%%%%%%%%%%%%%%%%%%%%%%%%%%%%%%%%%%%%%%%%%%%%%%%%%%%%%%%
%%%%%%%%%%%%%%%%%%%%%%%%%%%%%%%%%%%%%%%%%%%%%%%%%%%%%%%%%%%%%%%%%%%%%%%%%%%%%%%%%%%%%%%%%%%%%%%%%%%%

In cloud computing, there exist both latency-critical jobs and jobs that could tolerate different degrees of delay. The resource efficiency of a system is much dependent on the job's latency requirement. We propose a delay-differentiated pricing and service model where multiple SLAs are provided, as a complement to the existing on-demand service system. The structure of the market formed by the proposed model is studied and we thus derive the pricing rule under which the proposed framework forms a DSIC mechanism and the CSP's revenue is maximized. We consider two architectures for fulfilling SLAs: the first appears more prevalent and advanced in the literature while the second seems very simple. Our rigorous analysis discourages the adoption of the first architecture and supports the use of the second one. Finally, numerical results are given to show the viability of the proposed service model in comparison with a pure on-demand service system, showing a revenue improvement by up to 209.9\%.

\appendix

\section{Proofs}
\label{App-optimal-delay-Priority-queue}

%\vspace{0.25em}\noindent\textbf{Proof of Lemma~\ref{lemma-utility-difference-order}.}

%\subsection{Proof of Lemma~\ref{lemma-utility-difference-order}}

\vspace{0.25em}\noindent\textbf{Proof of Lemma~\ref{lemma-utility-difference-order}.} Let $\varphi\in [T, +\infty)$. It suffices to prove the conclusion that $g(\varphi)=u(\alpha_{2}, \varphi)-u(\alpha_{1}, \varphi)$ is an increasing function of $\varphi$; then, the lemma holds since $g(\varphi_{k_{2}})>g(\varphi_{k_{1}})$. To prove this, we note that the derivative of $g(\varphi)$ is
\begin{align*}
g^{\prime}(\varphi)=\frac{\partial{u(\alpha_{2}, \varphi)}}{\partial{\varphi}}-\frac{\partial{u(\alpha_{1}, \varphi)}}{\partial{\varphi}}.
\end{align*}
Since $\alpha_{1}>\alpha_{2}$, we have $g^{\prime}(\varphi)>0$ by the fourth point of Property~\ref{property-1}, and $g(\varphi)$ is increasing.

%\vspace{0.25em}\noindent\textbf{Proof of Lemma~\ref{lemma-sequential-choice}.}

%\subsection{Proof of Lemma~\ref{lemma-sequential-choice}}

\vspace{0.45em}\noindent\textbf{Proof of Lemma~\ref{lemma-sequential-choice}.} We prove this by contradiction. Suppose $k_{2} < k_{1}$ and the SLA delays satisfy $\varphi_{k_{2}}<\varphi_{k_{1}}$. The customer of type $\alpha_{1}$ (resp. $\alpha_{2}$) achieves the maximum surplus under the SLA $k_{1}$ (resp. $k_{2}$), and we thus have
\begin{align}
 u(\alpha_{1}, \varphi_{k_{1}})-p_{k_{1}} \geq u(\alpha_{1}, \varphi_{k_{2}})-p_{k_{2}} \label{ineq-1}\\
 u(\alpha_{2}, \varphi_{k_{1}})-p_{k_{1}} \leq u(\alpha_{2}, \varphi_{k_{2}})-p_{k_{2}} \label{ineq-2}
\end{align}
Multiplying (\ref{ineq-1}) by -1 and adding the resulting inequality to (\ref{ineq-2}), we have $u(\alpha_{2}, \varphi_{k_{1}})-u(\alpha_{1}, \varphi_{k_{1}}) \leq u(\alpha_{2}, \varphi_{k_{2}})-u(\alpha_{1}, \varphi_{k_{2}})$. However, since $\alpha_{1} > \alpha_{2}$ and $k_{2}<k_{1}$, we have by Lemma~\ref{lemma-utility-difference-order} that $u(\alpha_{1}, \varphi_{k_{2}})-u(\alpha_{1}, \varphi_{k_{1}}) > u(\alpha_{2}, \varphi_{k_{2}})-u(\alpha_{2}, \varphi_{k_{1}})$, which contradicts the previous inequality.

%\subsection{Proof of Lemma~\ref{theo-market-segmentation}}

\vspace{0.45em}\noindent\textbf{Proof of Lemma~\ref{theo-market-segmentation}.} Each type of customers will be assigned to some SLA, and $\Phi_{l}$ denotes the set of the types of the customers assigned to the $l$-th SLA for all $l\in [1, L]$. Let $\hat{\alpha}_{l}$ denote the maximum type in $\Phi_{l}$ such that only the customers of type $\alpha \leq \hat{\alpha}_{l}$ will possibly be assigned to the $l$-th SLA.
%By Lemma~\ref{lemma-sequential-choice}, a customer of larger type will be assigned to a SLA whose number is no larger than the SLA number of a customer of smaller type.
For all $l\in [1, L-1]$, when the customers of types $\hat{\alpha}_{l}$ and $\hat{\alpha}_{l+1}$ are respectively assigned the $l$-th and ($l+1$)-th SLAs, we have by Lemma~\ref{lemma-sequential-choice} that $\hat{\alpha}_{l}>\hat{\alpha}_{l+1}$, which can be easily proved by contradiction. A customer of type $\overline{\alpha}$ will be assigned to a SLA whose number is no larger than one (i.e., the first SLA) since $\overline{\alpha}\geq \hat{\alpha}_{1}$. Thus, we have $\hat{\alpha}_{1}=\overline{\alpha}$.

By Lemma~\ref{lemma-sequential-choice}, we also have that (\rmnum{1}) for all $l\in [1, L-1]$ every customer of type $\alpha\in \left(\hat{\alpha}_{l+1}, \hat{\alpha}_{l}\right]\cap\Phi$ will be assigned to a SLA whose number $l^{\prime}$ is no smaller than $l$ but no larger than $l+1$, and (\rmnum{2}) every customer of type $\alpha\in \left[\underline{\alpha}, \hat{\alpha}_{L}\right]\cap\Phi$ will be assigned to a SLA whose number is no smaller than $L$ since $\alpha\leq \hat{\alpha}_{L}$. In the first case, $\alpha> \hat{\alpha}_{l+1}$ and $\hat{\alpha}_{l+1}$ is the maximum type of $\Phi_{l+1}$; thus $l^{\prime}$ will be smaller than $l+1$ and equal $l$. The proposition thus holds.

%\subsection{Proof of Lemma~\ref{lemma-inequality}}

\vspace{0.45em}\noindent\textbf{Proof of Lemma~\ref{lemma-inequality}.} In the first case, if $\alpha=\hat{\alpha}_{l}$ and $l^{\prime}=l$, the surplus difference of the customer under the $l^{\prime}$-th and ($l^{\prime}-1$)-th SLAs is $\left(u(\hat{\alpha}_{l}, \varphi_{l})-p_{l}\right) - \left(u(\hat{\alpha}_{l}, \varphi_{l-1})-p_{l-1}\right)$; it equals zero due to Definition~\ref{def-optip}. Otherwise, we have either $\alpha<\hat{\alpha}_{l}$ or $l^{\prime}<l$: in the former, $\alpha < \hat{\alpha}_{l}\leq \hat{\alpha}_{l^{\prime}}$ since $l^{\prime}\in [2, l]$; in the latter, $\alpha \leq \hat{\alpha}_{l} < \hat{\alpha}_{l^{\prime}}$. Thus, we have $\alpha<\hat{\alpha}_{l^{\prime}}$. The surplus difference under two adjoining SLAs is
\begin{equation*}
\begin{split}
& \left(u(\alpha, \varphi_{l^{\prime}})-p_{l^{\prime}}\right) - \left(u(\alpha, \varphi_{l^{\prime}-1})-p_{l^{\prime}-1}\right) \\ \overset{(a)}{=} & \left( u\left(\hat{\alpha}_{l^{\prime}}, \varphi_{l^{\prime}-1}\right) - u\left(\hat{\alpha}_{l^{\prime}}, \varphi_{l^{\prime}}\right) \right) - \left(u(\alpha, \varphi_{l^{\prime}-1}) - u(\alpha, \varphi_{l^{\prime}})\right) \overset{(b)}{>} 0;
\end{split}
\end{equation*}
here, equation (a) is due to Definition~\ref{def-optip}, and (b) is due to Lemma~\ref{lemma-utility-difference-order}. In the second case, we have $\hat{\alpha}_{l^{\prime}+1}<\alpha$ since $\alpha\in (\hat{\alpha}_{l+1},\, \hat{\alpha}_{l}]$ and $l^{\prime}\geq l$, and the difference of the surpluses of the customer under the $l^{\prime}$-th and ($l^{\prime}+1$)-th SLAs is
\begin{equation*}
\begin{split}
& \left(u(\alpha, \varphi_{l^{\prime}})-p_{l^{\prime}}\right) - \left(u(\alpha, \varphi_{l^{\prime}+1})-p_{l^{\prime}+1}\right) \\ \overset{(c)}{=} & \left(u(\alpha, \varphi_{l^{\prime}})-u(\alpha, \varphi_{l^{\prime}+1})\right) - \left(u(\hat{\alpha}_{l^{\prime}+1}, \varphi_{l^{\prime}}) - u(\hat{\alpha}_{l^{\prime}+1}, \varphi_{l^{\prime}+1})\right)  \overset{(d)}{>} 0;
\end{split}
\end{equation*}
here, equation (c) is due to Definition~\ref{def-optip}, and (d) is due to Lemma~\ref{lemma-utility-difference-order}. Hence, the lemma holds.

%\subsection{Proof of Lemma~\ref{theo-SLA-assignment-maximum}}

\vspace{0.45em}\noindent\textbf{Proof of Lemma~\ref{theo-SLA-assignment-maximum}.} In the case that $\alpha\neq\hat{\alpha}_{l}$, we have by Lemma~\ref{lemma-inequality} the conclusion that, (\rmnum{1}) for all $l^{\prime}\in [2, l]$, the customer achieves a higher surplus under $l^{\prime}$-th SLA than under the ($l^{\prime}-1$)-th SLA, and (\rmnum{2}) for all $l^{\prime}\in [l, L-1]$, it achieves a higher surplus under the $l^{\prime}$-th SLA than under the ($l^{\prime}+1$)-th SLA; thus, the customer achieves the highest surplus under the $l$-th SLA. In the case that $\alpha=\hat{\alpha}_{l}$, we still have the above conclusion, except that the customer achieves the same surplus under the $l$-th and ($l-1$)-th SLAs when $l^{\prime}\in [2, l]$ and $l^{\prime}=l$. Hence, the customer achieves the maximum surplus under both the $l$-th and ($l-1$)-th SLAs. According to Definition~\ref{def-SLA-assignment}, the proposition holds in both cases.

%\subsection{Proof of Lemma~\ref{theo-sequence-1}}

\vspace{0.45em}\noindent\textbf{Proof of Lemma~\ref{theo-sequence-1}.} Let us consider a customer of type $\alpha\in\Phi_{l}$ who reports to the CSP that its type is $\alpha^{\prime}$. No matter what the other users do, we have by Proposition~\ref{theo-SLA-assignment-maximum} that it achieves the maximum surplus under the $l$-th SLA and will be assigned by the CSP to the $l$-th SLA when it truthfully reports its type, i.e., $\alpha^{\prime}=\alpha$. Thus, it cannot gain more by misreporting its type, since misreport can lead to that it is assigned to the $l$-th SLA or the other SLAs. The first point thus holds by Definition~\ref{def-dsic}.

The objective of our framework is to maximize (\ref{equa-objective-fun}); given the market segmentation $\hat{\alpha}_{1}, \hat{\alpha}_{2}, \cdots, \hat{\alpha}_{L+1}$ defined in Proposition~\ref{theo-market-segmentation}, the job arrival rate of each SLA is fixed by (\ref{equa-arrival-rate-l}) and we have the conclusion that the larger the SLA prices, the larger the value of $AG$. The first SLA's price $p_{1}$ is fixed and equals $p$. In order to guarantee the truthfulness of the customers of type $\alpha\in\Phi_{l}$, a necessary condition is that $u_{i_{l}}(\alpha, \varphi_{l-1}) - p_{l-1} \leq u_{i_{l}}(\alpha,\varphi_{l}) - p_{l}$, for all $l\in [2, L]$. Further, irrespective of the value of $p_{l-1}$, the maximum possible value of $p_{l}$ is $\hat{p}_{l}$ for all $l\in [2, L]$. Thus, the second point holds.

%\subsection{Proof of Lemma~\ref{proposi-optimal-delays}}

\vspace{0.45em}\noindent\textbf{Proof of Lemma~\ref{proposi-optimal-delays}.} We prove this by contradiction. We have $\varphi_{l}\geq t_{l}$ for all $l\in [2, L]$. Let us consider an optimal solution where the SLA delays and prices are $\varphi_{l}^{*}$ and $p_{l}^{*}$ for all $l\in [2, L]$, and the market segmentation is $\hat{\alpha}_{1}, \hat{\alpha}_{2}, \cdots, \hat{\alpha}_{L+1}$. Suppose there exists some SLA $l\in [2, L]$ such that $\varphi_{l}^{*}>t_{l}$; let $l^{\prime}$ denote the minimum such $l$, where $\varphi_{2}^{*}=t_{2}, \cdots, \varphi_{l^{\prime}-1}^{*}=t_{l^{\prime}-1}$ if $l^{\prime}> 2$. If we decrease the delay of the $l^{\prime}$-th SLA to $t_{l^{\prime}}$ and keep the others unchanged, we denote the corresponding prices by $\overline{p}_{1}, \cdots, \overline{p}_{L}$. It suffices to prove the conclusion that $\overline{p}_{l}>p_{l}^{*}$ for all $l\in [l^{\prime}, L]$ and $\overline{p}_{l}=p_{l}^{*}$ for all $l\in [2, l^{\prime}-1]$ if $l^{\prime}>2$. This will lead to that the revenue (\ref{equa-objective-fun}) increases, which contradicts the assumption that $p_{1}^{*}, \cdots, p_{L}^{*}$ are optimal; the proposition thus holds. Now, we prove the conclusion. The SLA prices are determined by Proposition~\ref{theo-sequence-1}. First, we have $p_{l}^{*}=\overline{p}_{l}$ for all $l\in [2, l^{\prime}-1]$ if $l^{\prime}> 2$; this is due to that $\varphi_{2}^{*}, \cdots, \varphi_{l^{\prime}-1}^{*}$ does not change. Second, for the $l^{\prime}$-th SLA, we have
\begin{align*}
\overline{p}_{l^{\prime}} & = \overline{p}_{l^{\prime}-1} + u(\hat{\alpha}_{l^{\prime}}, t_{l^{\prime}}) - u(\hat{\alpha}_{l^{\prime}}, t_{l^{\prime}-1})\\
&  \overset{(a)}{>} p_{l^{\prime}-1}^{*} + u(\hat{\alpha}_{l^{\prime}}, \varphi_{l^{\prime}}^{*}) - u(\hat{\alpha}_{l^{\prime}}, \varphi_{l^{\prime}-1}^{*})=p_{l^{\prime}}^{*}.
\end{align*}
The inequality (a) is due to that $\overline{p}_{l^{\prime}-1}=p_{l^{\prime}-1}^{*}$, $u(\hat{\alpha}_{l^{\prime}}, t_{l^{\prime}})>u(\hat{\alpha}_{l^{\prime}}, \varphi_{l^{\prime}}^{*})$, and $t_{l^{\prime}-1}=\varphi_{l^{\prime}-1}^{*}$. Third, for the ($l^{\prime}+1$)-th SLA, we have
\begin{equation*}
\begin{split}
& \overline{p}_{l^{\prime}+1} = \overline{p}_{l^{\prime}} + u(\hat{\alpha}_{l^{\prime}+1}, \varphi_{l^{\prime}+1}^{*}) - u(\hat{\alpha}_{l^{\prime}+1}, t_{l^{\prime}})\\
& = \overline{p}_{l^{\prime}-1} + u(\hat{\alpha}_{l^{\prime}}, t_{l^{\prime}}) - u(\hat{\alpha}_{l^{\prime}}, \varphi_{l^{\prime}-1}^{*}) + u(\hat{\alpha}_{l^{\prime}+1}, \varphi_{l^{\prime}+1}^{*}) - u(\hat{\alpha}_{l^{\prime}+1}, t_{l^{\prime}}) \\
& \overset{(b)}{>} p_{l^{\prime}-1}^{*} + u(\hat{\alpha}_{l^{\prime}}, \varphi_{l^{\prime}}^{*}) - u(\hat{\alpha}_{l^{\prime}}, \varphi_{l^{\prime}-1}^{*}) + u(\hat{\alpha}_{l^{\prime}+1}, \varphi_{l^{\prime}+1}^{*}) - u(\hat{\alpha}_{l^{\prime}+1}, \varphi_{l^{\prime}}) \\
& = p_{l^{\prime}+1}^{*}.
\end{split}
\end{equation*}
Here, the inequality (b) is due to Lemma~\ref{lemma-utility-difference-order}. Fourth, if $l^{\prime}+2\leq L$, for all $l\in [l^{\prime}+2, L]$, we have by a simple mathematical induction that
\begin{align*}
\overline{p}_{l} & =\overline{p}_{l-1} + u(\hat{\alpha}_{l}, \varphi_{l}^{*}) - u(\hat{\alpha}_{l}, \varphi_{l-1}^{*})\\
& \overset{(c)}{>} p_{l-1}^{*} + u(\hat{\alpha}_{l}, \varphi_{l}^{*}) - u(\hat{\alpha}_{l}, \varphi_{l-1}^{*})=p_{l^{\prime}}^{*}.
\end{align*}
Here, the inequality (c) is due to $\overline{p}_{l-1}>p_{l-1}^{*}$.

\vspace{0.45em}\noindent\textbf{Proof of Proposition~\ref{proposi-opti-configuration}.} Algorithm~\ref{Algo-OptiDelay-separated} searches each possible pair of $(\alpha_{1}, \alpha_{2}, \cdots$, $\alpha_{L+1})$ and $(i_{1}, i_{2}, \cdots, i_{L+1})$ respectively in $\mathcal{A}$ and $\mathcal{M}$ (lines 1, 2, 3, 14, 4, 5, 13 of Algorithm~\ref{Algo-OptiDelay-separated}), and computes the corresponding revenue under this pair (lines 6-10). Among all pairs that have been searched so far, it records the current maximum revenue and the corresponding SLA delays and prices, and the numbers of servers assigned to SLAs (lines 1, 11, 12). Thus, the algorithm will return the optimal solution. The sizes of $\mathcal{M}$ and $\mathcal{A}$ are respectively polynomial in $m$ and $n$ (i.e., $\binom{m}{L-1}$ and $\binom{n}{L-1}$). The loop in line 4 is nested in the loop in line 2; hence, the time complexity is $\mathcal{O}\left(m^{L-1}\cdot n^{L-1}\right)$.

\section{Additional Experiments}

%\subsection{{\color{blue}Comparison with a Variant of \cite{Abhishek12,Dierks19}}}

As seen in Section~\ref{sec.Introduction}, our framework differs from \cite{Abhishek12,Dierks19} in several aspects. Nevertheless, the service model in Section~3 and 4 is generic. The architecture of \cite{Abhishek12,Dierks19} can be adapted to our model, and roughly viewed as a hybrid of the PBS and SMS architectures. Specifically, all servers are separated into two parts: the first are used to fulfill the first SLA, as done by the first module of the SMS architecture; the second use priority queues to fulfill the SLAs $2, \cdots, L$, as done by the PBS architecture. Specially, when the number of SLAs is two (i.e., $L=2$), the SMS and hybrid architectures are the same and the model has the same performance under both architectures, which can achieve a significantly larger revenue than the pure on-demand service model. Generally, the PBS-based service system has a performance close to the on-demand service system but performs worse than the SMS-based system since the PBS architecture achieves a lower utilization. It can be expected that the hybrid architecture has a in-between performance, as shown later.

\begin{figure}[t]
  \centering
  \includegraphics[width=2.75in]{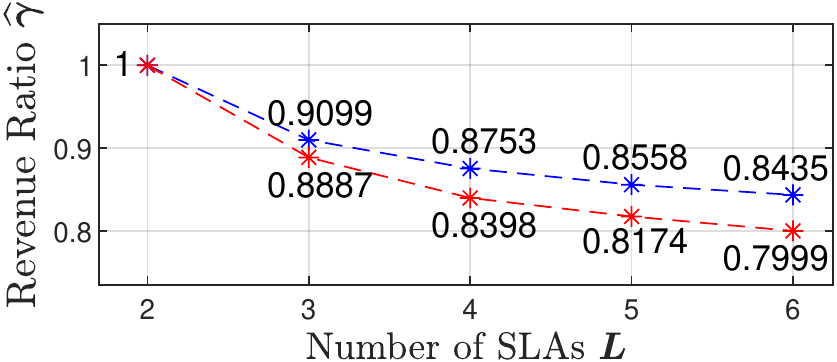}
  \caption{\textbf{Revenue Ratio $\hat{\gamma}$ with $L$ SLAs}: the red (resp. blue) stars correspond to the case of low (resp. high) delay-tolerance.}\label{Fig-Revenue-Ratio}
\end{figure}

We denote by $G_{hyb}^{\ast}$ the maximum revenue achieved by our service model under the hybrid architecture. For all $l\in [2, L]$, let $\hat{\lambda}_{l}^{\prime}$ denote the total job arrival rate of SLAs $2, \cdots, l$ at a single server; we can derive the actual delay $t_{l}$ of the $l$-th SLA by (\ref{equa-exp}) and the equation (\ref{equa-waiting-time-random-0}) for the PBS architecture, and have
\begin{align}\label{equa-waiting-time-random-Hybrid}
t_{l} = \hat{\lambda}_{L}^{\prime}/((1-\hat{\lambda}_{l-1}^{\prime})\cdot(1-\hat{\lambda}_{l}^{\prime})),
\end{align}
where $\hat{\lambda}_{1}^{\prime}$ is set to zero trivially. The value of $G_{hyb}^{\ast}$ can be computed by a small modification of the line 7 of Algorithm~\ref{Algo-OptiDelay-separated} where for all $l\in [2, L]$ we change to use (\ref{equa-waiting-time-random-Hybrid}) to compute $t_{l}$. The revenue ratio $\hat{\gamma}$, defined below, is used to show which of the SMS and Hybrid architectures is better: $$\hat{\gamma}=G_{hyb}^{\ast}/G_{sms}^{\ast}.$$ If $\hat{\gamma}\leq 1$, the service model under the hybrid architecture will be no better than the SMS-based service system. This is exactly shown by the numerical results illustrated in Fig.~\ref{Fig-Revenue-Ratio}.

The reason for $\hat{\gamma}\leq 1$ is mainly due to (\rmnum{1}) the correlation of the SLA delays in the hybrid architecture and (\rmnum{2}) the delay-sensitivity of the jobs of SLA 2, both of which place limitations on the power of some jobs with larger delay-tolerance to achieve a higher utilization of servers. First, we have by (\ref{equa-waiting-time-random-Hybrid}) that the actual delays $t_{2}, \cdots, t_{L}$ are all constrained by the total job arrival rate $\hat{\lambda}_{L}^{\prime}$, which is also the average load per server in the second part. Second, there is a sequence $i_{1}$, $i_{2}$, $\cdots$, $i_{L}$ for mapping jobs to SLAs, as described in the last subsubsection. In the second part, the most delay-sensitive jobs have a type $\hat{\alpha}_{2}=1/\varphi_{0,i_{2}}^{\prime}$, and are assigned to the second SLA, which requires a small SLA delay $\varphi_{2}$ to guarantee that the SLA price $p_{2}$ does not decrease to a negligible value. This further leads to a small $\hat{\lambda}_{L}^{\prime}$.

For example, in the low delay-tolerance case with $L=4$, the first and second parts have 51 and 49 servers respectively. The market segmentation is $(i_{2}, i_{3}, i_{4})=(13, 19, 30)$ and we correspondingly have $(\varphi_{0,i_{2}}, \varphi_{0,i_{3}}, \varphi_{0,i_{4}})=(0.29, 0.41, 0.63)$.
%$\varphi_{0,2}=0.29$, $\varphi_{0,3}=0.41$, and $\varphi_{0,4}=0.63$.
The SLA delays and prices are as follows: $(\varphi_{1}, \varphi_{2}, \varphi_{3}, \varphi_{4})=(0.05, 0.1590, 0.1709, 0.1973)$ and $(p_{1}, p_{2}, p_{3}, p_{4})=(1, 0.9063, 0.8963, 0.8889)$.
%$\varphi_{2}=0.1590$, $\varphi_{3}=0.1709$, $\varphi_{4}=0.1973$, and $p_{2}=0.9063$, $p_{3}=0.8963$, $p_{4}=0.8889$.
Specially, $\varphi_{2}$ has to be small and is around 0.5 times $\varphi_{0,i_{2}}$ to guarantee that the price $p_{2}=u(\hat{\alpha}_{2}, \varphi_{2})$ is not low, as introduced in Section~\ref{sec.delay-cost-curve}. By (\ref{equa-waiting-time-random-Hybrid}), the value of $\varphi_{2}$ further limits that $\hat{\lambda}_{L}^{\prime}$ has to be small where $\varphi_{2}=t_{2}$. In the experiments, we have $(\hat{\lambda}_{2}^{\prime}, \hat{\lambda}_{3}^{\prime}, \hat{\lambda}_{4}^{\prime})=(0.02449, 0.06939, 0.1551)$.
%$\hat{\lambda}_{2}^{\prime}=0.02449$, $\hat{\lambda}_{3}^{\prime}=0.06939$, and $\hat{\lambda}_{4}^{\prime}= 0.1551$.
This leads to that the second part of servers achieve relatively low utilization and revenue. Finally, for the $m$ servers, the average load per server is 0.1.

In contrast, the delays of different SLAs in the SMS architecture are independent by (\ref{equa-waiting-time-separated-random}), which unlocks the power of trading the job's delay-tolerance for a higher utilization. Specifically, the numbers of servers assigned to different SLAs are $(m_{1}, m_{2}, m_{3}, m_{4})=(21, 24, 28, 27)$.
%$m_{1}=21$, $m_{2}=24$, $m_{3}=28$, and $m_{4}=27$.
The market segmentation is $(i_{2}, i_{3}, i_{4})=(5, 12, 26)$ and correspondingly $(\varphi_{0,i_{2}}, \varphi_{0,i_{3}}, \varphi_{0,i_{4}})=(0.13, 0.27, 0.55)$.
%$\varphi_{0,2}=0.13$, $\varphi_{0,3}=0.27$, and $\varphi_{0,4}=0.55$.
The SLA delays and prices are $(\varphi_{1}, \varphi_{2}, \varphi_{3}, \varphi_{4})=(0.05, 0.07527, 0.1364, 0.2857)$ and $(p_{1}, p_{2}, p_{3}, p_{4})=(1, 0.9685, 0.9095, 0.8099)$.
%$\varphi_{2}=0.07527$, $\varphi_{3}=0.1364$ and $\varphi_{4}=0.2857$, and the SLA prices are $p_{2}=0.9685$, $p_{3}=0.9095$ and $p_{4}=0.0.8099$.
Although the value of $\varphi_{2}$ is still small, it imposes no constraints on the value of $\lambda_{4}$, i.e., the average load per server of the fourth SLA. In the experiments, we have $(\lambda_{2}, \lambda_{3}, \lambda_{4})=(0.07000, 0.1200, 0.2222)$.
%$\lambda_{2}=0.07000$, $\lambda_{3}=0.1200$, and $\lambda_{4}= 0.2222$.
For the $m$ servers, the average load per server is 0.12, which is larger than the one in the hybrid architecture, and the revenue ratio $\hat{\gamma}$=0.8753.

\end{document}